\documentclass[useAMS,usenatbib]{mn2e}
\usepackage{times,graphicx,epsfig}

\catcode`\"=\active\let"=\"
\font\cc cmcsc10
\def\3{\ss }
\def\I{{\cal I}}
\def\J{{\cal J}}
\def\D{{\cal D}}
\def\twosqpi{{2 \over \sqrt{\pi}}}
\def\intl{\int\limits}
\def\Erf{{\rm Erf}}
\def\vesc{v_{{\rm esc}}}
\def\vescr{v_{{\rm esc,r}}}
\def\vesct{v_{{\rm esc,t}}}

\def\erf{{\rm erf}}
\def\nbody#1{{\cc Nbody#1}{}}
\def\plusplus{\raise 0.3ex\hbox{${\scriptstyle ++}$}{}}
 \def\dedet#1#2{\left(\delta #1\over \delta t\right)_{\rm #2}}

\title[Anisotropic gaseous models of tidally limited star clusters ]
{Anisotropic gaseous models of tidally limited star clusters -- \\
 comparison with other methods}
\author[R. Spurzem, M. Giersz, K. Takahashi, A. Ernst]
{R. Spurzem$^1$\thanks{Rhine-Stellar Dynamics Network (RSDN) {\tt http://www.ari.uni-heidelberg.de/rsdn}},
M. Giersz$^2$, K.~Takahashi$^3$, A.~Ernst$^1$\footnotemark[1]           \\
$^1$Astronomisches Rechen-Institut, Zentr. Astron. Univ. Heidelberg (ZAH), M"onchhofstrasse 12-14,
   69120 Heidelberg, Germany \\
$^2$ Nicolaus Copernicus Astronomical Center (CAMK),
   Polish Academy of Science, ul. Bartycka 18, 00-716 Warsaw, Poland\\
$^3$Department of Astronomy, University of Tokyo, 7-3-1 Hongo, Bunkyo-ku,
Tokyo 113-0033, Japan}

\begin{document}

\maketitle

\begin{abstract}
We present new models of the evolution and dissolution of
star clusters evolving under the combined influence of internal relaxation and
external tidal fields, using the anisotropic gaseous model based on the
Fokker-Planck approximation, and a new escaper loss cone model. This
model borrows ideas from loss cones of stellar distributions near massive
black holes, and describes physical processes related to escaping stars
by a simple model based on two timescales and a diffusion process.
We compare our results with those of direct $N$-body models
and of direct numerical solutions of the orbit-averaged
Fokker-Planck equation.
For this comparative study we limit ourselves to idealized single
point mass star clusters, in order to present a detailed
study of the physical processes determining the rate of
mass loss, core collapse and other features of the
system's evolution. With the positive results of our study the
path is now open in the future to use the computationally efficient gaseous models
for future studies with more realism (mass spectrum, stellar evolution).
\end{abstract}


\section{Introduction}
 Dynamical modelling of globular clusters and other collisional
 stellar systems (like galactic nuclei, rich open clusters, and
 rich galaxy clusters) still poses a considerable
 challenge for both theory and computational requirements
 (in hardware and software). On the theoretical side
 the validity of certain assumptions
 used in statistical modelling based on the Fokker-Planck
 (henceforth FP) and other approximations is poorly known.
 Stochastic noise in a discrete $N$-body system
 and the impossibility to directly model realistic particle numbers
 with the presently available hardware, are a considerable
 challenge for the computational side.

 Detailed comparisons of the results obtained with the
 different methods for single mass isolated star
 clusters have been performed (Giersz \&
 Heggie 1994a,b, Giersz \& Spurzem 1994,
 Spurzem \& Aarseth 1996, Spurzem 1996). They include
 theoretical models such as the
 direct numerical solution of the orbit-averaged
 1D FP equation for isotropic systems (Cohn 1980),
 isotropic (Heggie 1984) and
 anisotropic gaseous models (henceforth AGM) (Louis \& Spurzem 1991, Spurzem 1994)
 and direct $N$-body simulations using standard $N$-body codes
 (\nbody{5}, Aarseth 1985, Spurzem \& Aarseth 1996; \nbody{2},
 Makino \& Aarseth 1992; {\nbody{4},} Makino 1996; \nbody{6\plusplus},
 Spurzem 1999, Aarseth 1999a,b, 2003).  All the cited work, however,
 only dealt with idealized single-mass models. There are very
 few attempts yet to extend the quantitative comparisons to
 more realistic star clusters containing different mass bins or
 even a continuous mass spectrum (Spurzem \& Takahashi 1995,
 Giersz \& Heggie 1996, G{\" u}rkan, Freitag \& Rasio 2004).

 On the side of the FP models there have been two major
 recent developments. Takahashi (1995, 1996, 1997) has published
 new FP models for spherically symmetric star clusters,
 based on the numerical solution of the orbit-averaged 2D FP
 equation (solving the FP
 equation for the distribution $f=f(E,J^2)$ as a function of
 energy and angular momentum, on an $(E,J^2)$-mesh).
 Drukier et al. (1999) have published
 results from another 2D FP code based on the original
 Cohn (1979) code. In such 2D FP models
 anisotropy, i.e. the possible difference between radial
 and tangential velocity dispersions in spherical clusters is taken
 into account. Although the late, self-similar stages of core collapse
 are not affected very much by anisotropy (Louis \& Spurzem 1991),
 intermediate and outer zones of globular clusters, say outside roughly
 the Lagrangian radius containing 30 \% of the total mass, do exhibit
 fair amounts of anisotropy, in theoretical model simulations as well
 as according to parameterized model fits (Lupton, Gunn \& Griffin 1987).
 In contrast to AGMs the 2D FP
 models contain less inherent model approximations; they do not
 assume a certain form of the heat conductivity and closure relations
 between the third order moments as in the case of AGM.
 Furthermore, the latter contains a numerical
 constant $\lambda $ (Spurzem 1996), which is of order unity,
 but its numerical value has to be determined from comparisons with
 proper FP or $N$-body models.

 Secondly, another 2D FP model has been worked out recently
 for the case of axisymmetric rotating star clusters (Einsel \& Spurzem
 1999, Kim et al. 2002, Kim, Lee \& Spurzem 2004, Fiestas, Spurzem \& Kim 2005).
 Here, the distribution function is assumed to be a function
 of energy $E$ and the $z$-component of angular momentum $J_z$ only;
 a possible dependence of the distribution function on a third integral
 is neglected. As in the spherically symmetric case the neglecting
 of an integral of motion is equivalent to the assumption
 of isotropy, here between the velocity dispersions in the meridional
 plane ($r$ and $z$ directions); anisotropy between that velocity
dispersion
 and that in the equatorial plane ($\phi$-direction), however, is
included.

 Thirdly, there is an elegant alternative way to generate models of
 star clusters, which can correctly reproduce the stochastic features
 of real star clusters, but without really integrating
 all orbits directly as in an $N$-body simulation. They rely on the FP
 approximation and (hitherto) spherical  symmetry, but their data
 structure is very similar to an $N$-body model. These so-called Monte Carlo
 models were recently redeveloped by Giersz (1996, 1998, 2001), and by Rasio
 and collaborators
 (Joshi, Rasio \& Portegies Zwart 2000,
 Watters, Joshi \& Rasio 2000, Joshi, Nave \& Rasio 2001,
 Fregeau et al. 2003, G\"urkan, Freitag \& Rasio 2004). For
 another approach, reviving H\'enon's superstar method, compare the work
 by Freitag (Freitag 2000, Freitag \& Benz 2001, 2002).
 The basic idea is  to have pseudo-particles, whose
 orbital parameters are given in a smooth, self-consistent potential.
 However, their orbital motion is not explicitly followed; to
 model interactions with other particles like two-body relaxation
 by distant encounters or strong interactions between binaries and
 field stars, a position of the particle in its orbit and further
 free parameters of the individual encounter are picked from
 an appropriate distribution by using  random numbers.
 A hybrid variant of the Monte Carlo technique combined with a gaseous model has been
 proposed by Spurzem \& Giersz (1996), and applied to systems with a
 large number of primordial binaries by Giersz \& Spurzem (2000) and
 Giersz \& Spurzem (2003). The hybrid method uses a Monte Carlo model
 for binaries or any other object for which a statistical description,
 as used by the gaseous model, is not appropriate, due to small numbers
 of objects or unknown analytic cross sections for interaction processes.
 The method is particularly useful for investigating evolution of large
 stellar systems with realistic fraction of primordial binaries, but
 could also be used in future to include the build-up of massive stars
 and blue stragglers by stellar collisions, for example.

 In the present and near future a wealth of detailed data on globular clusters
 will become available by e.g. the Hubble Space Telescope and the new 8m-class
 terrestrial telescopes such as Gemini and the Very Large Telescope (VLT),
 for extragalactic as well as Milky Way clusters. These data cover
 luminosity functions and derived mass functions,
 color-magnitude diagrams, population and kinematical analysis,
 including binaries and compact stellar evolution remnants,
 detailed two-dimensional
 proper motion and radial velocity data, and tidal tails spanning over arcs
 several degrees wide (Koch et al. 2004).
 With detailed observational data such as from King at al. (1998),  Piotto \& Zocalli (1999),
 Rubenstein \& Bailyn (1999), Ibata et al. (1999), Piotto et al. (1999),
 Grillmair et al. (1999), Shara et al. (1998), Odenkirchen et al. (2001),
 Hansen et al. (2002), Richer et al. (2002) to mention only the few
 recently appeared papers), an easily reproducible reliable
 modelling becomes more important than before. For that purpose
 a few more ingredients are urgently required in the models in
 addition to anisotropy and rotation: a
 mass spectrum, a tidal field, and the influence of stellar evolution on
 the dynamical evolution of the cluster.

 While it is easy in principle to include all these in a direct $N$-body
 simulation, and considerable effort goes in the construction of new
 hardware and software for that purpose (Hut \& Makino 1999, Makino et
al. 1997, Makino \& Taiji 1998,
 Aarseth 1999a,b, 2003), the life span of globular clusters extends over
 tens to hundreds of thousands of crossing times, taken at the half-mass
radius,
 which requires a high accuracy direct $N$-body code for its modelling.
 Despite the enormous advances in hardware and software efficiency,
 still modelling, say, of a few hundred thousand particles is very tedious.
 For parameter studies one has
 to rely on lower particle numbers and prescriptions to scale the
 results to larger $N$ (Aarseth \& Heggie 1998, Baumgardt 2001). So we still need
 the fast but approximate theoretical models. Takahashi,
 Lee \& Inagaki (1997) published first results for anisotropic clusters
 in a tidal field,
 and Takahashi \& Lee (2000) extended their study to multimass
 clusters.
 Takahashi \& Portegies Zwart (1998, 2000), Portegies Zwart \& Takahashi (1999) compared
 the influence of stellar evolutionary mass loss as measured in
 $N$-body and FP models. However, still the tidal boundary
 poses an unsolved problem in the pure point mass case. Another
 difficulty is that $N$-body models and theory are
 difficult to match, since an energy cutoff as it is usual in the
 FP models has not yet been applied
 in $N$-body simulations (with the exception of one of the models
 kindly provided by E. Kim for our comparisons, see below); usually
 direct $N$-body models employ a tidal
 radius cutoff.

 Moreover, in the theoretical models one has to decide which criteria to
use
 for a star to escape. In contrast to isolated clusters energy and
angular
 momentum are not exact integrals of motion in a tidal field, and one
should
 consider more appropriate quantities like the Jacobian.
 However, even then it is possible that a star satisfying an
 escape criterion stays for many orbital times close to the cluster
 and even can be scattered back to the cluster (Baumgardt 2001), so
 from an observers viewpoint they do not escape immediately; indeed
Fukushige \& Heggie (2000) propose a new form of an energy dependent
escape time scale, which is formally infinite for stars at the tidal
energy (note a
 similar problem in relation to dwarf spheroidal galaxies in tidal
 fields raised by Kroupa 1998 and Klessen \& Kroupa 1998).

 In this situation we propose to look at simple cases and simplified
 models in more detail first. For example it appears to be useful to
 separate the dynamical effects of stellar evolution from those induced
 by the interplay of internal relaxation and a tidal field, and to look
 at a single mass case first. A gaseous model of a star cluster (here used
 an anisotropic gaseous model or AGM, see {\bf http://www.ari.uni-heidelberg.de/gaseous-model})
 is a very
 simple tool with which to understand, on the basis of an idealized
 model, the physical processes
 acting on star clusters. It has been very successfully used to
 detect gravothermal oscillations (Bettwieser \& Sugimoto 1984) and
 to discuss effects present in multi-mass star clusters (Spurzem \&
 Takahashi
 1995) and in systems with primordial binaries (Heggie \& Aarseth 1992).

 Thus in this paper we present a new approach to include a tidal field
 into AGM of star clusters, which has not been
 tackled before. The processes of escape and relaxation are treated in a
 simplified parameterized way and compared to models using an orbit averaged
 Fokker-Planck equation and direct $N$-body simulations. Also the different
 parameters of our escaper model are varied to understand their influence
 on the results.
 We will demonstrate that the
 gaseous model helps in the physical understanding of the tidal
 escape process and allows a deeper insight into the processes going on
 in other models.  With this the way is paved for an application of this
 same gaseous model including more realistic properties of star clusters,
 such as a mass spectrum, stellar evolution and primordial binaries.

\section{The models}

The anisotropic gaseous model (AGM) is based on local moments
of the stellar velocity distribution function. From the
local Fokker-Planck equation (Rosenbluth, McDonald \& Judd 1957)
moment equations up to second order are taken and closed in
third order by a phenomenological heat flux equation
(Lynden-Bell \& Eggleton 1980, Bettwieser 1983, Heggie 1984,
Louis \& Spurzem 1991). Diffusion coefficients are determined
self-consistently, including the anisotropy of the background
distribution of scatterers, but locally without any orbit
average (Spurzem \& Takahashi 1995). Two free parameters
($\lambda$ scaling the heat flux, and $\lambda_A$ scaling
the collisional decay of anisotropy) are determined by
quantitative comparison with orbit-averaged Fokker-Planck
and direct $N$-body models, a standard binary heating term due to
three-body binary energy generation is used; this and the
presently used form of the closure and all equations can be
found in Giersz \& Spurzem (1994). Note that our present
variant (sometimes called a one-flux closure, see Louis 1990
and Louis \& Spurzem 1991) is the only one, which behaves
reasonably well for the case of multi-mass systems with
equipartition and dynamical friction, as was shown in
Spurzem \& Takahashi (1995).
We present here an enhancement to the AGM method to handle an external
tidal field, and also use data of the 2D FP method worked out by Takahashi
(1995, 1996, 1997) and $N$-body data kindly provided by S. Deiters
and D.C. Heggie and E. Kim (2003) for comparison purposes.

The equations of AGM are solved numerically by an implicit Henyey
method as described in Giersz \& Spurzem (1994) or Spurzem (1996).
Collisional terms were evaluated locally with self-consistent
anisotropic test and background star distributions (Giersz \&
Spurzem 1994, Spurzem \& Takahashi 1995). The energy generation due
to binaries is the same as in the FP models, though in the gaseous
models it is applied locally, not in an orbit-averaged way.

To describe the process of stars escaping from a cluster in a tidal
field we adopt simple approximations which are outlined in the
following. In the spherically symmetric AGM full phase space
information is reduced to the knowledge of moments of the velocity
distribution up to third order (density, bulk radial mass motion,
radial and tangential velocity dispersion, radial fluxes of radial
and tangential energy, see e.g. Spurzem 1994 or Giersz \& Spurzem
1994 for details of the models). Since the mass and energy fluxes
drive the quasistatic evolution of the system under relaxation
timescales, to first order the most significant moments shaping the
velocity distribution function are density and the radial and
tangential velocity dispersions.

In the standard tidal cutoff picture a star is considered to be an
escaper if its integrals of motion (that of an isolated cluster),
energy and angular momentum, fulfil certain criteria related to the
tidal radius $r_t$ or tidal cutoff energy $E_t$ of the cluster
(Takahashi, Lee \& Inagaki 1997). We approximate the real
distribution function of the tidally limited cluster in the gaseous
model as follows: a tidally unlimited local anisotropic
Schwarzschild Boltzmann distribution
\begin{eqnarray}
\lefteqn{f(r,v_r,v_\theta,v_\phi) =} \nonumber \\
 & & {\rho\over(2\pi)^{3/2}\sigma_r\sigma_t^2}
 \exp\Bigl(-{(v_r-u)^2\over{2\sigma_r^2}} -
 {(v_\theta^2 + v_\phi^2)\over{2\sigma_t^2}}\Bigr)
\label{1}
\end{eqnarray}
(where $\rho$ is the mass density, $v_r$, $v_{\theta}$, $v_{\phi}$ are the individual stellar
velocities in a local Cartesian coordinate system, whose axes are
tangential to the radial, polar and azimuthal directions on a
sphere, $u$ is the bulk mass transport velocity, $\sigma_r$,
$\sigma_t$ are the radial and tangential velocity dispersion,
respectively) is used to compute the fraction of stars $X_e$ which
would be in the escaper region if a tidal limit would be imposed on
such a cluster by
\begin{equation}
X_e = \int\limits_e f d^3 v\Bigr/ \int\limits f d^3 v
\label{2}
\end{equation}
Here, the index $e$ at the integral denotes an integration over escape velocity
space; for example in case of an energy criterion we have
\begin{equation}
v_r^2 + v_\theta^2 + v_\phi^2 > \vesc^2 \equiv 2(\Phi_t-\Phi(r))
\label{3}
\end{equation}
as escape condition with the potential $\Phi_t$ given at the tidal
radius ($r_t = (M/3M_G)^{1/3}R_G$, where $M$ is the total cluster
mass, $M_G$ is the parent galaxy mass and $R_G$ is the distance
between the galaxy and cluster centers), $v_{esc}$ is the escape
velocity and $\Phi(r)$ is the potential at the distance $r$. For an
apocentre criterion, taking into account energy and angular momentum
conservation (at distance $r$ and apocentre distance set to $r_t$),
it is straightforward to obtain from Eq.~\ref{3},
\begin{equation}
{v_r^2\over \vescr^2} + {(v_\theta^2 + v_\phi^2) \over \vesct^2 } > 1
\label{4}
\end{equation}
as a condition for escape, where we have used the convenient
definitions $\vescr \equiv \vesc $ and $\vesct^2 \equiv \vesc^2\cdot
r_t^2/(r_t^2-r^2)$. The integral in Eq. \ref{2} is a 3D-integral
over an unbounded domain with an ellipsoidal boundary. In order to
solve it we first integrate for the non-escaping fraction of stars
$X_{ne}$ (which are inside the ellipsoidal boundary), which is a
bound integration domain, and due to the normalization of the
velocity distribution function we can determine $X_e = 1 - X_{ne}$.
Fortunately, the integral for $X_{ne}$ can be solved analytically by
using error functions and some similar kind of integrals, including
Dawson's integral, for which a routine can be obtained from
Numerical Recipes (Press et al. 1986). In the Appendix some more
detail of this derivation is given. Here, we show the result, using
abbreviations:
\begin{equation}
a \equiv {\vescr \over \sqrt{2}\sigma_r } \ \ b \equiv {\vesct \over \sqrt{2}\sigma_t }
\label{4a}
\end{equation}
and getting the end result

\begin{equation}
X_e = 1 - \erf(a) + \cases{ \displaystyle a \exp(-b^2)
{\left(\erf(G)\over G \right)}   & $a > b$ \cr
\displaystyle a
\exp(-b^2) {\I(H)\over H }   & $a < b$ }
\end{equation}
with the definitions of $G^2 \equiv a^2-b^2$ and $H^2 \equiv b^2-a^2$. The special
function $\I(x)$ is related to Dawson's integral and defined in the
Appendix. Similarly to $X_e$ we also compute the fraction of energy of
the stellar system (radial and tangential) belonging to the escaper space by
\begin{equation}
X_r = \int\limits_e v_r^2 f d^3 v\Bigr/
         \int\limits v_r^2 f d^3 v
\label{6}
\end{equation}
\begin{equation}
X_t = \int\limits_e (v_\theta^2+v_\phi^2) f d^3 v \Bigr/
         \int\limits (v_\theta^2+v_\phi^2) f d^3 v \ .
\label{7}
\end{equation}
If $X_r = X_t = X_e$ all escaping stars would have the same average specific
energy as the non-escaping ones. Generally this should not be the case,
since escaping stars would tend to have higher specific energies,
in such case the difference between
$X_r$, $X_t$ and $X_e$ tells us something about the specific energy
of the escapers as compared to the non-escaping stars. Again the
values are determined by first integrating over the bounded domain
of the non-escaping stars and getting the complement due to the
normalization. Our results from the Appendix are:
\begin{eqnarray}
 X_r \!&=&\! 1\! -\! 2\, \Erf(a)\! + \!\cases{
\displaystyle{2a^3 \exp(-b^2) {\Erf(G)\over G^3 } }  & $a > b$ \cr
    \displaystyle{  2a^3 \exp(-b^2) {\J(H)\over H^3 }}   & $a < b$ } \\
 X_t \!&=&\! 1\! -\! \erf(a)\! +\!
 \cases{ \displaystyle{ a (1+b^2) \exp(-b^2) {\erf(G)\over G }}  & \cr
                        \ \ \   \displaystyle{- ab^2 \exp(-b^2) {\Erf(G)\over G^3 }}   \!\! & $a > b$ \cr
                                \displaystyle{a (1+b^2) \exp(-b^2) {\I(G)\over G }}   & \cr\smallskip
                        \ \ \   \displaystyle{- ab^2 \exp(-b^2) {\J(H)\over H^3 } }  \!\! & $a < b$ }
\label{8}
\end{eqnarray}
where $\erf$, $\Erf$, $\I$, and $\J$ are the error function and
specially defined generalizations of it (see Appendix).

In a realistic case, however, the distribution function will be different
from a Schwarzschild-Boltzmann function and the escaper fraction of
velocity space will be populated by a few stars only, which are on their
way out to leave the cluster. Therefore, we assume that the density of stars
$\rho_e$ prone to escape in reality is smaller than $X_e\rho$ by a factor
$k<1$, which will be referred as to filling factor. Hence we have
$\rho_e \equiv kX_e\rho$, and for the radial and tangential energies
$p_{re} \equiv kX_rp_r$, $p_{te} \equiv kX_tp_t$, where $p_r = \rho\sigma_r^2$ and $p_t = \rho\sigma_t^2$.
Such procedure can be seen in close connection to
the ansatz of King's models (King 1966), which just use a lowered Maxwellian to
model the distribution function of a tidally limited cluster.
Then our ansatz for mass and energy loss of the cluster is
\begin{equation}
\dedet{\rho}{e} = - {\rho_e \over \alpha t_{\rm cross}}
\label{10}
\end{equation}
\begin{equation}
\dedet{p_r}{e} = -{p_{re} \over \alpha t_{\rm cross}}
\label{11}
\end{equation}
\begin{equation}
\dedet{p_t}{e} = -{p_{te} \over \alpha t_{\rm cross}}
\label{12}
\end{equation}
where $t_{\rm cross}$ denotes a crossing time to reach the tidal radius
with the radial escape velocity, and $\alpha$ is a free parameter
with which one can describe the unknown process of removal of escaping
stars from the cluster.

To complete our model the time evolution of the filling factor $k$
has to be described. We are doing this in a close analogy to the
loss cone description of Frank \& Rees (1976) and Amaro-Seoane, Freitag \& Spurzem (2004) for stars to be
swallowed by a central black hole. The process which brings stars
into the escaper region is two-body relaxation, so to the first order
we think that the timescale $t_{\rm in}$ to refill the ``loss
cone'' (which is here the escaper region of velocity space) is
assumed to be
\begin{equation}
t_{\rm in} \equiv \beta t_{\rm rx},
\label{13}
\end{equation}
where $t_{\rm rx}$ is the local relaxation time.
We keep a free parameter $\beta$ because some details of the process,
e.g. to what extent it is a true diffusion process, remain unclear at
the moment. The timescale for stars to leave the loss-cone is
\begin{equation}
t_{\rm out} \equiv \alpha t_{\rm cross}.
\label{14}
\end{equation}
Hence we have at each radius $r$ an
approximate ``diffusion'' equation describing how stars enter and leave
the escaper region of velocity space:
\begin{equation}
{d\rho_e\over dt}=
 -{\rho_e \over t_{\rm out}}+
 {(1 - k)\rho X_e\over t_{\rm in}}
\label{18}
\end{equation}
We put the term ``diffusion'' for this process in quotation marks
here, because the underlying physical process is angular momentum
diffusion in stellar systems, which is properly described only in a
2D Fokker-Planck model, using proper diffusion coefficients. The
diffusion process tends to establish isotropy, i.e. an equal
distribution of angular momenta - distribution function does not
depend on angular momentum across the loss-cone and in its vicinity
in velocity space. Here, we model this process in very simplified way
by the ansatz that the diffusion term is proportional to the
difference in densities inside (low angular momentum) and outside
(high angular momentum) the loss cone. So, if $k=1$, we have an
isotropic equilibrium and the term vanishes, if $k<1$ the diffusion
refills the loss cone with a rate proportional to $1/t_{\rm in}$,
where $t_{\rm in}$ is the standard local relaxation time times a
factor $\beta$ of order unity. The reader is also referred to
Amaro-Seoane, Freitag, \& Spurzem (2004), where a similar concept is
used for star accretion onto a massive central black hole.

Suppose, we have to readjust the filling factor $k$ of escaper space
for some model at time $t$ and radial shell $r$ during the numerical
solution of the gaseous model equations. Then we can consider
locally $\rho$ and $X_e$ as constant, and find after dividing a factor
$\rho X_e$ out of the above equation
\begin{equation}
{dk\over dt} = - {k \over t_{\rm out}} + {(1-k)\over t_{\rm in}}
\label{15}
\end{equation}
It can be solved directly only if $t_{\rm in}$ and $t_{\rm out}$ are
held fixed. But we use this solution only to advance $k(t)$ from
one time step to another, then reinitialising it with new
values of $t_{\rm in}$ and $t_{\rm out}$. If the timestep is
small enough such that $t_{\rm in}$ and $t_{\rm out}$ do not
change much this is a good approximation.
\begin{eqnarray}
k(t)&=&k_0\exp\Bigl(-{K_0(t-t_0)\over t_{\rm out}}\Bigr) + \nonumber \\
 & & {t_{\rm out}\over t_{\rm in}K_0}\cdot
 \biggl(1-\exp\Bigl(-{K_0(t-t_0)\over t_{\rm out}}\Bigr)\biggr)
\label{19}
\end{eqnarray}
Here $k_0 \equiv k(t_0)$ and $K_0 \equiv 1+t_{\rm out}/t_{\rm in}$; for
$t\rightarrow\infty$ we find a stationary solution
$k_\infty \equiv 1/(1+t_{\rm in}/t_{\rm out})$. In our numerical models we
compute the new filling factor $k=k(r)$ according to Eq.~\ref{19} in
every radial shell between the time steps of the Henyey iteration by
setting $k_0 \equiv k(t_0)$ and $t \equiv t_0 + dt$ with the model time step $dt$.
For small time steps a linear approximation to the exponential
function would be sufficient, which would again lead to linearly
discretised form of Eq.~\ref{15}; since the computational costs are
small compared to the Henyey iteration we prefer to be on the safe
side and use the full exponential function expression Eq.~\ref{19}
for the recomputation of $k$.

Note, some
special meanings of the parameters $\alpha$ and $\beta$. Choosing for
example a very small $\alpha \ll 1$ is equivalent to an immediate
removal of escaping stars from the system, as e.g. Chernoff \& Weinberg
(1990) and other FP models usually treat the escapers. If
$\alpha $ is of the order of one it means that we allow for some time
before the actual removal of the stars in a complex tidal gravitational
field really takes place. If $\beta \ll 1$ the loss cone region is
very quickly refilled (practically all the time is full), and if $\beta \gg 1$
the loss cone is not refilled.

This simplified diffusion and escape model, described just by the
two time-scales $t_{\rm out}$ and $t_{\rm in}$, coming with the
corresponding two parameters $\alpha$ and $\beta$, is being used
only in those gaseous model shells were $E<E_t$, i.e. the entire
shell is still bound to the system and its total energy is smaller
than the tidal energy. For shells whose energy as a whole is
lifted above the tidal energy $E_t$ we follow a prescription
originally proposed by Lee \& Ostriker (1987, LO87). They point
out that stars at the tidal energy need very long time to actually
escape from the cluster, and only if their energy is higher than
that, they will asymptotically escape with a time scale
proportional to the crossing time at the tidal radius. The ansatz
of LO87 for the evolution of the phase space distribution function
$f=f(E,J^2)$ is
\begin{equation}
{ \partial f \over \partial t} = - \alpha_{\rm FP} f
 \Bigl[1 - \Bigl({\vert E\vert \over \vert E_t\vert }\Bigr)^3\Bigr]^{1/2} \cdot
 {1\over 2\pi} \sqrt{{4\pi\over 3} G \rho_{\rm av}} \ .
\label{16}
\end{equation}
Note that we have added here the modulus of energies $E$, $E_t$
because our sign convention for the energy ($E<0$) is different
from that of LO87. $\rho_{\rm av}$ denotes the average density of
the cluster, so the last term is actually inversely
proportional to a crossing time at the tidal radius.
 Portegies Zwart \& Takahashi (1999) find that such a model provides
 a good match between FP models and direct $N$-body
 simulation. According to their choice of words, there
 is no ``crisis'' in Fokker-Planck models due to
 the $N$-dependance of the dissolution time,
 provided the parameter $\alpha_{\rm FP}$ is adjusted to
 a value of order unity. LO87 varied $\alpha_{\rm FP}$ from
 0.2 to 5.0, while Takahashi \& Portegies Zwart (2000) in
 an extensive multi-mass study for realistic globular clusters
 (with mass spectrum and stellar evolution) claim that
 $\alpha_{\rm FP}=2.5$ is the best value. We will discuss
 the role of $\alpha_{\rm FP}$ in our models presented here later;
 physically $\alpha_{\rm FP}$ brings in another time scale of
 mass loss (rather than the relaxation time), so its role is
 most prominent for small particle numbers, where both time
 scales are comparable. If $N$ is larger, typically 32000 or
 more, the crossing time is small compared to the relaxation
 time, so the latter completely determines the overall evolution
 of the system. But if in the course of the tidal evolution
 of a star cluster the particle number drops, at some moment
 the terms of LO87 will become important, and so the
 time of final dissolution of any cluster will depend on $\alpha_{\rm FP}$.

 In the anisotropic gaseous model we implement the dynamical
 mass loss of LO87 simply by applying a mass loss term in
 the densities corresponding to Eq.\ref{16}, if in a radial
 shell there is $E>E_t$:
 \begin{equation}
 { \partial \rho \over \partial t} = -\alpha_{\rm FP} \rho
  \Bigl[1 - \Bigl({\vert E\vert \over \vert E_t\vert }\Bigr)^3\Bigr]^{1/2} \cdot
 {1\over 2\pi} \sqrt{{4\pi\over 3} G \rho_{\rm av}}\ .
 \label{17}
 \end{equation}
 The tidal energy $E_t$ is determined in the gaseous model
 simply by $E_t = GM(t)/r_t(t)$, where $r_t$ results from the
 standard condition that the average density of the entire
 cluster remains constant, and $M(t)$ is the time-dependent
 total mass ($r_t(t) = (M(t)/M_0)^{1/3}r_t(0)$, where $M_0$ and
 $r_t(0)$ are the initial total mass and tidal radius, respectively).

\begin{table}
 \caption{Parameters of the initial models}
 \label{t1}
 \begin{tabular}{@{}ccccccc}
  Name& N & $W_0$ & $\alpha$ & $\beta$ & $\alpha_{FP}$ & k - trh \\
  \hline
  AGM-1 & 128000 & 3 & 1 & 1 & 1 & No - No \\
  AGM-2 & 128000 & 6 & 1 & 1 & 1 & No - No \\
  AGM-3 & 128000 & 9 & 1 & 1 & 1 & No - No \\
  \hline
  AGM-4 & 64000 & 3 & 1 & 1 & 1 & No - No \\
  AGM-5 & 64000 & 6 & 1 & 1 & 1 & No - No \\
  AGM-6 & 64000 & 9 & 1 & 1 & 1 & No - No \\
  \hline
  AGM-7 & 32000 & 3 & 1 & 1 & 1 & No - No \\
  AGM-8 & 32000 & 6 & 1 & 1 & 1 & No - No \\
  AGM-9 & 32000 & 6 & 1 & 0.5 & 1 & No - Yes \\
  AGM-10 & 32000 & 6 & 2.5 & 0.5 & 1 & No - Yes \\
  AGM-11 & 32000 & 6 & 5 & 0.5 & 1 & No - Yes \\
  AGM-12 & 32000 & 6 & 1 & 1 & 1 & No - Yes \\
  AGM-13 & 32000 & 6 & 2.5 & 1 & 1 & No - Yes \\
  AGM-14 & 32000 & 6 & 5 & 1 & 1 & No - Yes \\
  AGM-15 & 32000 & 6 & 1 & 2 & 1 & No - Yes \\
  AGM-16 & 32000 & 6 & 2.5 & 2 & 1 & No -Yes \\
  AGM-17 & 32000 & 6 & 5 & 2 & 1 & No - Yes \\
  AGM-18 & 32000 & 6 & 1 & 1 & 0 & No - No \\
  AGM-19 & 32000 & 6 & 1 & 1 & 0.2 & No - No \\
  AGM-20 & 32000 & 6 & 1 & 1 & 2.5 & No - No \\
  AGM-21 & 32000 & 6 & 1 & 1 & 5 & No - No \\
  AGM-22 & 32000 & 6 & 1 & 0.5 & 1 & Yes - No \\
  AGM-23 & 32000 & 6 & 1 & 1 & 1 & Yes - No \\
  AGM-24 & 32000 & 6 & 1 & 2 & 1 & Yes - No\\
  AGM-25 & 32000 & 9 & 1 & 1 & 1 & No - No \\
  \hline
  AGM-26 & 16000 & 3 & 1 & 1 & 1 & No - No \\
  AGM-27 & 16000 & 6 & 1 & 1 & 1 & No - No \\
  AGM-28 & 16000 & 6 & 1 & 2 & 1 & No - No \\
  AGM-29 & 16000 & 9 & 1 & 1 & 1 & No - No \\
  \hline
  AGM-30 & 5000 & 3 & 1 & 1 & 1 & No - No\\
  AGM-31 & 5000 & 6 & 1 & 1 & 1 & No - No \\
  AGM-32 & 5000 & 6 & 1 & 1 & 0 & No - No \\
  AGM-33 & 5000 & 6 & 1 & 1 & 2.5 & No - No \\
  AGM-34 & 5000 & 6 & 1 & 1 & 1 & Yes - No \\
  AGM-35 & 5000 & 9 & 1 & 1 & 1 & No - No\\
  \hline
  AGM-36 & 1000 & 3 & 1 & 1 & 1 & No - No\\
  AGM-37 & 1000 & 6 & 1 & 1 & 1 & No - No \\
  AGM-38 & 1000 & 9 & 1 & 1 & 1 & No - No\\
  \hline
  IFP-1 & 32000 & 3 & 1 & 1 & 1 & No - No \\
  IFP-2 & 32000 & 6 & 1 & 1 & 1 & No - No \\
  IFP-3 & 1000 & 3 & 1 & 1 & 1 & No - No \\
  IFP-4 & 1000 & 6 & 1 & 1 & 1 & No - No \\
  \hline
  AFP-1 & 5000 & 6 & 1 & 1 & 1 & No - No \\
  \hline
  Nbody-1 & 16000 & 6 &  &  &  &  \\
  Nbody-2 & 5000 & 6 &  &  &  &  \\
  \hline
  \end{tabular}

\medskip
 {AGM - anisotropic gaseous models, IFP - isotropic Fokker-Planck models,
 AFP - anisotropic Fokker-Planck models,
 Nbody - $N$-body models. $N$ - initial number of stars in the model,
 $W_0$ - concentration parameter of
 the King model, $\alpha$ and $\beta$ - parameters describing the
 process of removal of escaping stars and replenishing the escape
 region, respectively (see text - discussion of Eqs. 11 - 16),
 $\alpha_{\rm FP}$ - parameter of Eq. 20 (see text), k - full or
 in equilibrium loss cone of escaping stars (No - equilibrium,
 Yes - full), trh - initial or actual half mass relaxation time
 used in diffusion equation (No - initial, Yes - actual) - see
 text.}
 \end{table}
 At any time step we redetermine the tidal radius from the
 present mass of the cluster, and remove all shells which
 fall outside of the newly determined tidal radius. This
 leads typically to a small zone where $E>E_t$, but still $r<r_t$
 in which the LO87 dynamical mass loss procedure applies.
 Sometimes, in the very late phases of rapid mass loss
 before dissolution it can happen that shells inside
 the tidal radius have even positive energy ($E>0$). If this
 happens the shell is immediately removed, as if it would
 lie outside the tidal radius.

 Our model of mass loss in a tidal field can be determined
 by in total three parameters: $\alpha$ and $\beta$, which
 describe the time scales of the loss cone description
 for stars entering and leaving the escaper loss cone,
 and $\alpha_{\rm FP}$, which scales dynamical mass loss
 terms at the outer boundary.

 Loss terms according to Eqs.~\ref{10}, \ref{11}, \ref{12} and \ref{17} are
 applied as additional terms to the gaseous model equations
 during the Henyey iteration. It turns out that both the
 standard loss terms of Eqs.~\ref{10}, \ref{11}, \ref{12} as well as the
 LO87 term Eq.~\ref{17} are necessary to get a reasonable
 fit of mass loss in a cluster in a tidal field as compared
 with direct $N$-body as well as FP models. The
 filling degree $k(r)$ can be either set to a constant
 initial value of $k=0$ or $k=1$ according to empty or full
 loss cones, or the stationary value $k_\infty$ can be
 used which generally depends on $r$. For most models discussed in
 this paper we used the stationary value $k_\infty$ initially, see Table 1.
 It is interesting to note that in an unpublished study of
 Vicary (1997) using a Lagrangian gaseous model (Heggie 1984)
 the same qualitative behaviour of mass loss was found as here
 using a much simpler description based upon the loss of mass
 shells across the tidal boundary only. Unfortunately
 further details of that model are not available now.

\section{Initial Setup}

All models discussed in the paper are described by idealized single
mass star clusters under the influence of external tidal field. The
initial positions and velocities of all stars were drawn from a King
model.  The set of initial King models were characterized by $W_0 =
3$, $6$ and $9$, number of stars by $N = 1000$, $5000$, $16000$,
$32000$, $64000$ and $128000$. The range of parameters $\alpha$,
$\beta$ and $\alpha_{\rm FP}$ are $1, 2.5, 5$; $0.5, 1, 2$ and $0,
0.2, 0.5, 1, 2.5, 5$, respectively. The most natural values for
parameters are: $\alpha = 1$ $\beta = 1$ and $\alpha_{\rm FP} = 1$,
which means that the time scales for the processes characterized by
these parameters are described exactly by the local crossing and
relaxation times. The initial models are described in Table 1. The
data for anisotropic Fokker-Planck model (AFP) and $N$-body model
for particle number $N = 5000$ were kindly provided by Kim (2003),
and $N$-body data for $N = 16000$ by Heggie and Deiters.

To properly describe, in AGM, the process of mass removal connected
with loss cone effect and removal of unbounded shells it was
necessary to increase the spatial resolution of the model,
particularly close to the tidal boundary. Removal of stars or shells
from the gaseous model is a big challenge for the numerical
algorithm, when the boundary of the model is closed. Enhanced mass
loss generates errors of density, velocity, energy and mass
distributions, which lead to uncontrolled error in the total mass
and energy of the system. To reduce the error to the acceptable
level (a few percent) we decided to use, instead of the
logarithmically equidistant grid-points, the mesh, which resolution
was enhanced towards the tidal boundary. After some numerical
experiments the number of shells was chosen to: 541, 783, 967 for
$W_0 = 3, 6, 9$, respectively. This guarantee that mass and energy
errors was always less than $7\%$. Unfortunately, the code became
less efficient. The time of calculations is linearly proportional to
the number of shells.

For the computational units the standard $N$-body units (Heggie \&
Mathieu 1986) were used: total mass $M = 1$, $G = 1$ and initial
total energy of the cluster equal to $-1/4$. In all presented in
this paper figures the unit of time is expressed in terms of the
initial half-mass relaxation time, which for single mass system and
$N$-body units is equal to (Spitzer 1987):
\begin{equation}
t_{rh0} = {0.138 {N_0r_{h0}^{3/2}} \over \ln{(\gamma N_0)}},
\end{equation}
where $N_0$, $r_{h0}$ are initial number of stars and half-mass radius, respectively. The
value of coefficient in the Coulomb logarithm is taken to be $\gamma = 0.11$ (Giersz
and Heggie 1994a).

\section{Results}

Fig. 1a shows the effect of varying $\beta$ on the evolution of
the central density in the gaseous model with $N = 32000$ particles
in the case of constant $\alpha = 1$. First, one can see that the
evolution of the system already in pre-collapse varies for different
$\beta$; the reason is that the smaller $\beta$ the quicker
refilling of the loss cone region and the higher mass loss from the
system. In other words, if stars are removed from the cluster at a
certain multiple of the crossing time, this process has a different
speed relative to relaxation for each $N$. With constant $\alpha $
Fig. 1b shows, that a variation of $\beta$ just increases or
decreases the efficiency of the escape process in a constant amount
for each particle number. The differences in the dissolution time
are even greater than in the collapse time. This is connected with
the fact that the dynamical mass loss described by Eq. 20 is most
prominent for the small particle numbers - models with smaller
$\beta$ and higher mass loss.

\begin{figure}
\vspace{1.0 cm}
\psfig{figure=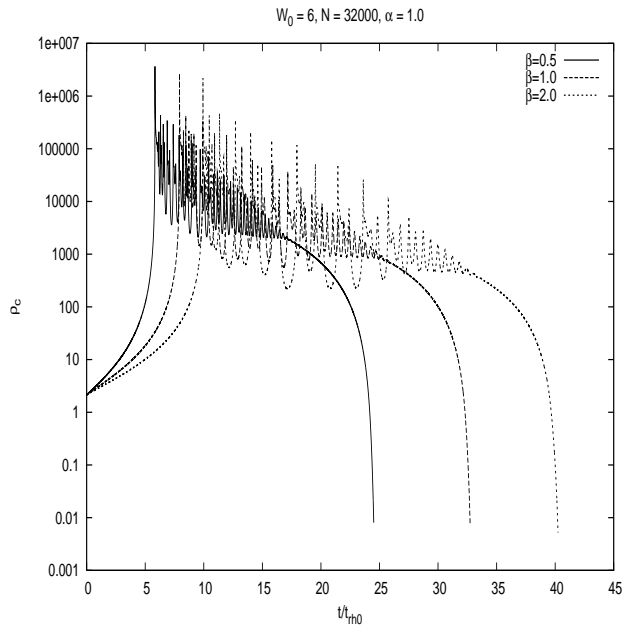, height=0.475\textwidth, width=0.475\textwidth, angle=-90}
\caption{{\bf a:} Central density of initial King model $W_0=6$ as
a function of time for gaseous models with 32000 particles,
constant $\alpha=1.0$ and varying $\beta = 0.5,1.0,2.0$ as
indicated in the key. Increasing $\beta$ leads monotonically to
increasing lifetimes of the cluster, because the mass loss
 time scale is larger.}
 \end{figure}

 \begin{figure}
 \vspace{1.0 cm}
\psfig{figure=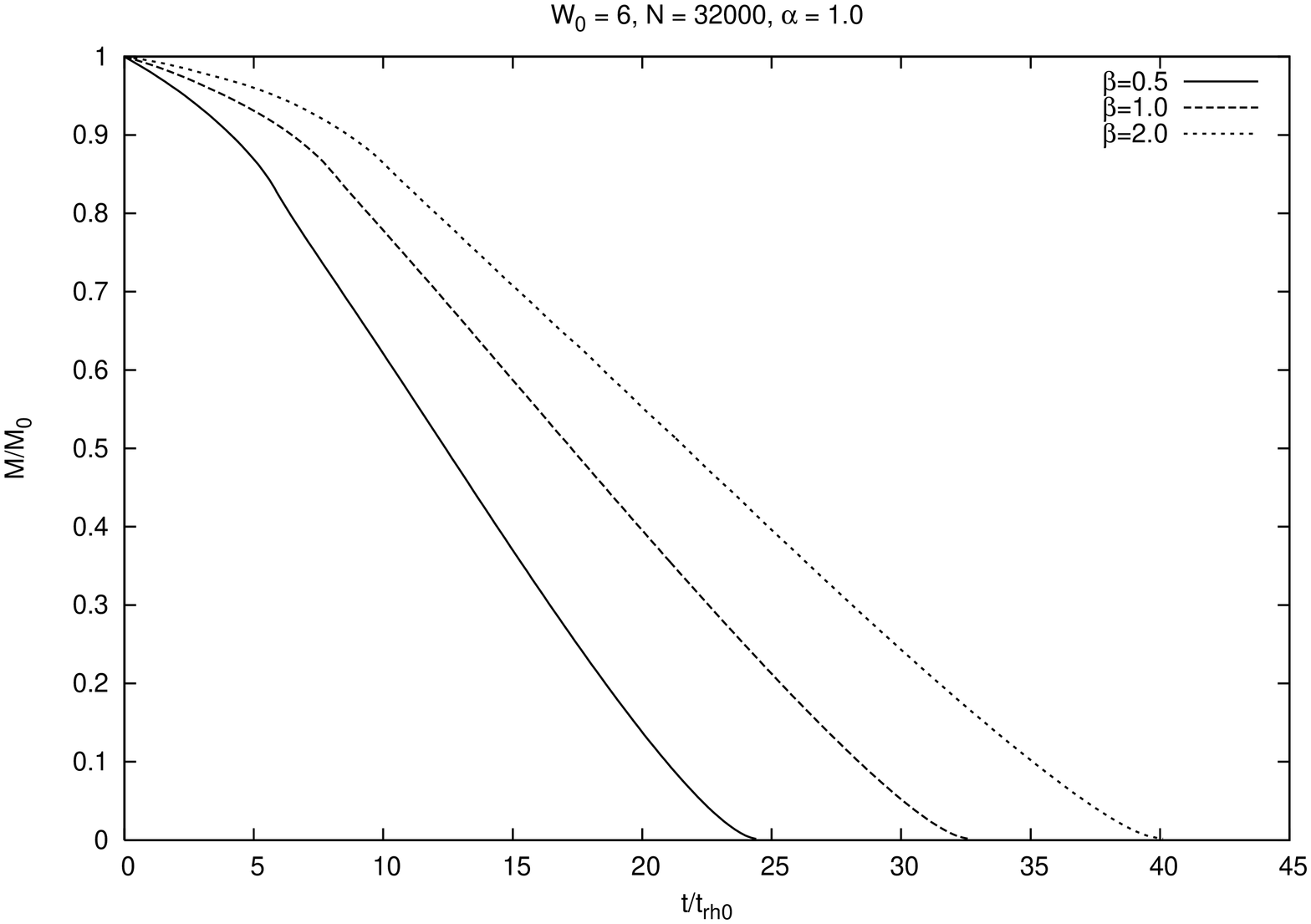,height=0.475\textwidth, width=0.475\textwidth, angle=-90}
 \contcaption{{\bf b:} As Fig. 1a, but for the total mass of the system.}
 \label{f1}
 \end{figure}

 In contrast to that Figs. 2 show the effect
 of varying $\alpha$ with constant
 $\beta$. There is almost no effect of changing $\alpha$ on
 collapse, and a small effect on dissolution times. The rate
 of mass loss is only slightly faster for small $\alpha$ than for larger
 $\alpha$ (see Fig. 2b). This is consistent with the picture that the rate of
 mass loss strongly depends on the rate of refilling the loss cone region. Star
 first have to be scattered by the relaxation process to the loss cone region and
 then on the crossing time scale escape from the system.

 \begin{figure}
 \vspace{1.0 cm}
\psfig{figure=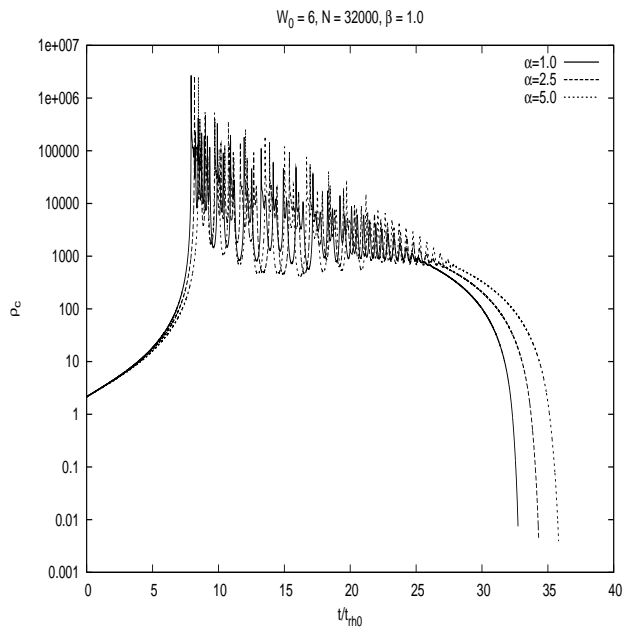,height=0.475\textwidth, width=0.475\textwidth, angle=-90}
 \caption{{\bf a:} Central density of initial King model $W_0=6$ as
 a function of time for gaseous models with 32000 particles,
 constant $\beta = 1.0$ and varying $\alpha = 1.0, 2.5, 5.0$ as indicated in the key.}
 \end{figure}

 \begin{figure}
 \vspace{1.0 cm}
\psfig{figure=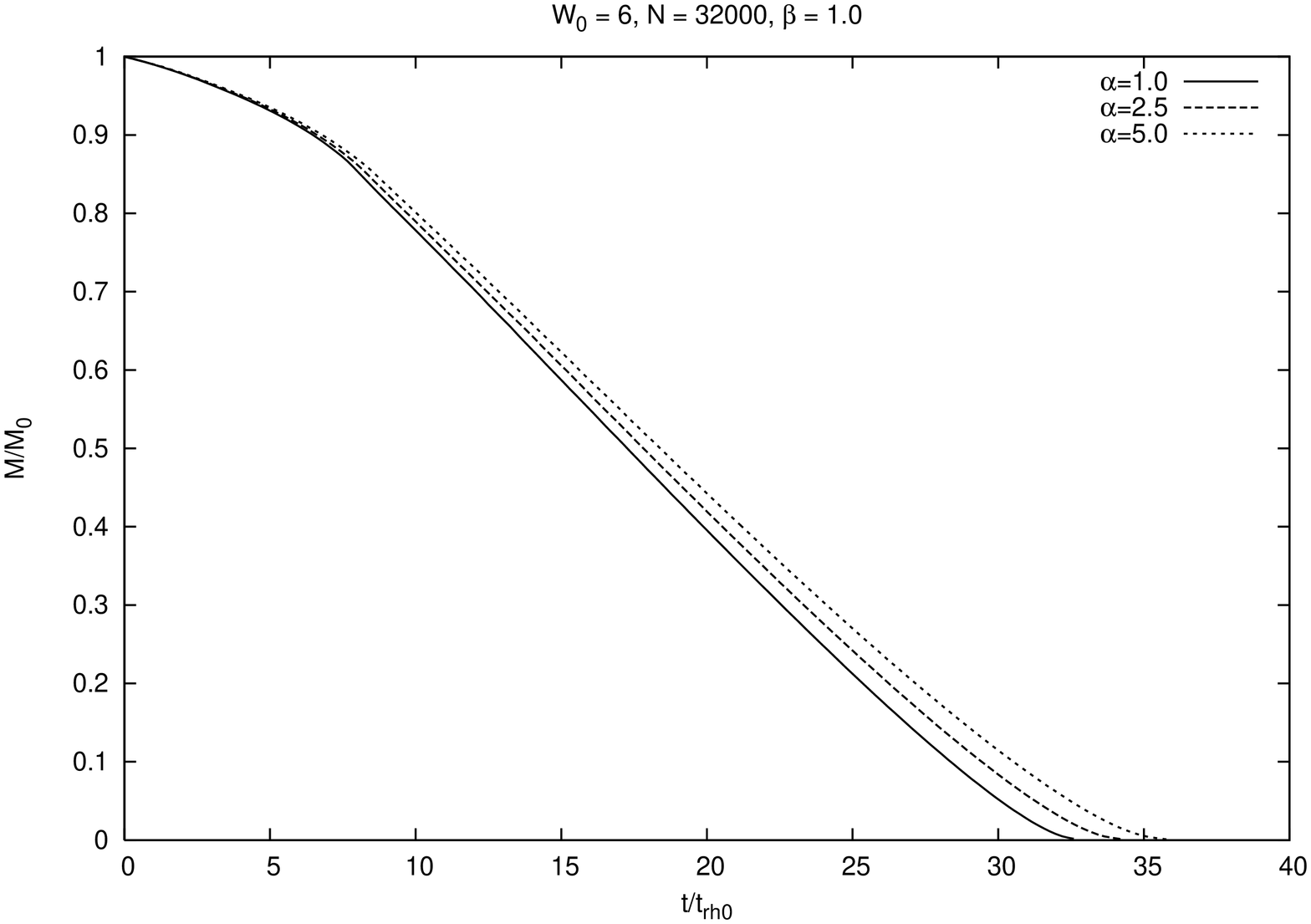,height=0.475\textwidth, width=0.475\textwidth, angle=-90}
 \contcaption{{\bf b:} As Fig. 2a, but for the total mass of the system.}
 \label{f2}
 \end{figure}

 \begin{figure}
 \vspace{1.0 cm}
\psfig{figure=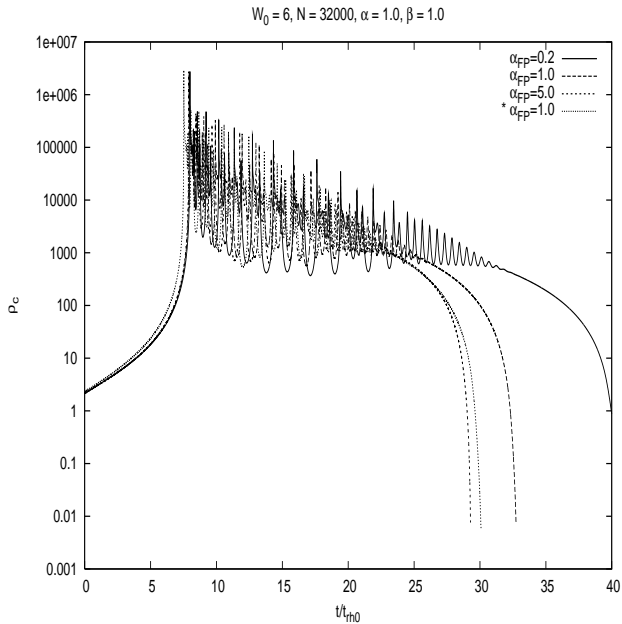,height=0.475\textwidth, width=0.475\textwidth, angle=-90}
 \caption{{\bf a:} Central density of initial King model $W_0=6$ as
 a function of time for gaseous models with 32000 particles,
 constant $\beta = 1.0$ and $\alpha = 1.0$. Here we vary
 the dynamical mass loss constant $\alpha_{\rm FP}$ as indicated in
 the key. The * symbol denotes a case started with full loss cones, $k = 1$ (see text).}
 \end{figure}

 \begin{figure}
 \vspace{1.0 cm}
\psfig{figure=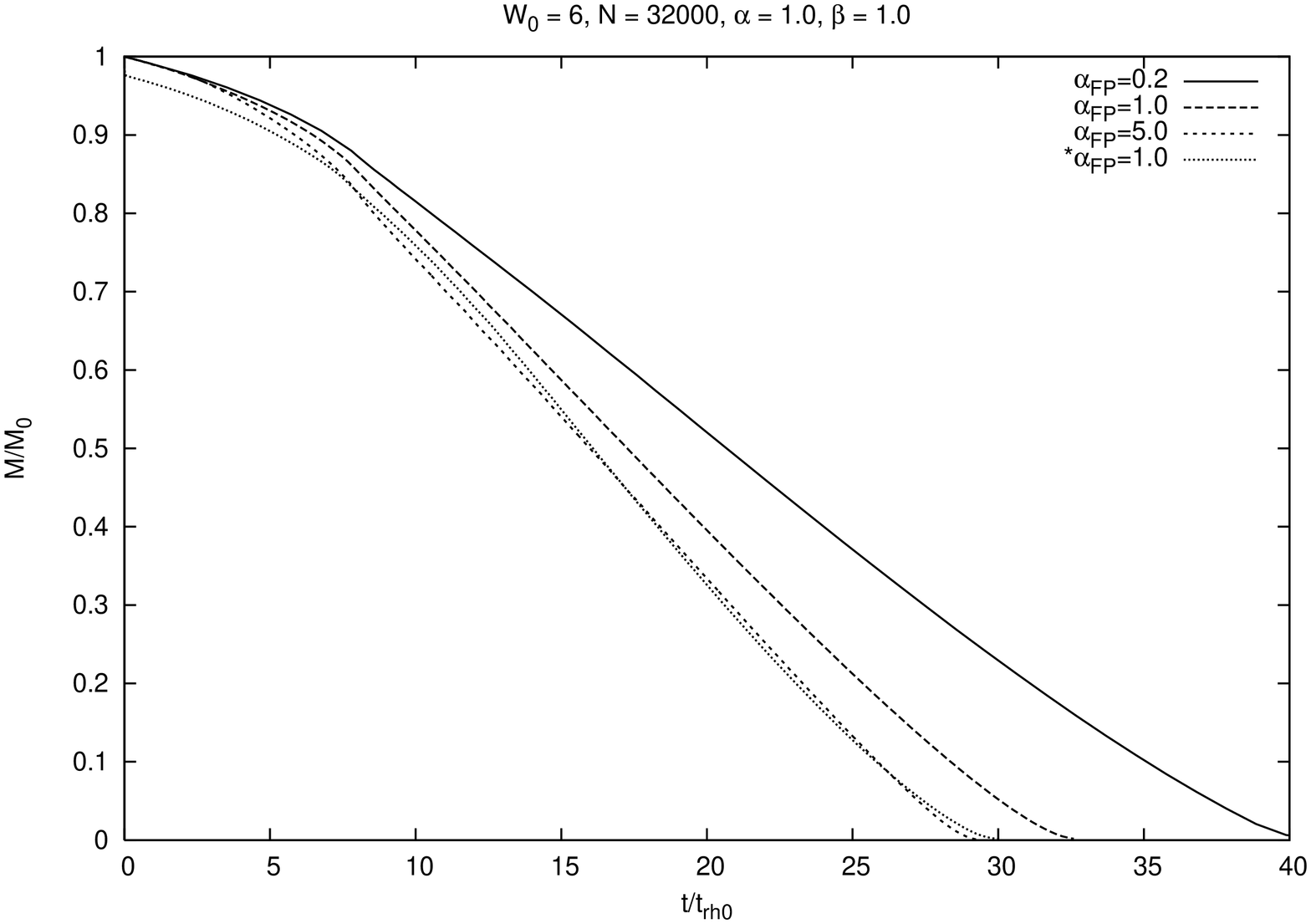,height=0.475\textwidth, width=0.475\textwidth, angle=-90}
 \contcaption{{\bf b:} As Fig. 3a, but for the total mass of the system.}
 \label{f2esc}
 \end{figure}

 From Figs. 1 and 2 one can see that the diffusion time scale $t_{in} = \beta t_{rx}$,
 i.e. the parameter $\beta$ has a significant influence on the core collapse time
 and the final dissolution time, by changing the mass loss rate. If the mass loss
 rate is larger, the core collapse accelerates, since the actual half-mass relaxation
 time becomes smaller than the initial one, and obviously the final dissolution time
 is different. We have used $\alpha_{\rm FP}=1.0$ for all models, unless stated otherwise.

 Figs. 3 demonstrate the role of $\alpha_{\rm FP}$ -- it hardly changes the
 early evolution, however, the onset of the final dramatic dissolution phase
 can be influenced by it. In the extreme case, where
 $\alpha_{\rm FP} \rightarrow 0$ the final dissolution is taking extremely long time.
 This is clearly an unphysical case, as has already been stated by LO87 and
 Takahashi \& Portegies Zwart (1998, 2000), Portegies Zwart \& Takahashi (1999).
 We also show one example, where we have started initially with
 full loss cones, i.e. the phase space part in the escaper region is
 fully populated for all regions of the system. It leads to a quick mass loss
 in the beginning, due to draining of the full loss cone in a dynamical time,
 which cannot be resolved in the figure, and thereafter we end up with a
 somewhat faster evolution. In all other models we start with a stationary
 filling degree of the loss cone, $k_{\infty}$.

 Now, we turn to comparison of our models with direct $N$-body results.
 There are only few models available, and most of them are not published, or
 only partly published. We use here one 16000 $N$-body simulation
 kindly provided by S. Deiters and D.C. Heggie, for $W_0=6$, and
 another one, using 5000 particles, kindly provided by E. Kim (2003).

 Comparing these model results with those of our AGM
 (Figs. 4)  we find that the natural choice of $\alpha=1.0$,
 $\beta=1.0$ and $\alpha_{\rm FP}=1.0$
 provides a fairly good match of the mass loss between $N$-body and
 AGM for both particle numbers. However, the core collapse times
 differ by a non-negligible amount. The increase
 of the parameter $\beta$ to values much larger then one will not solve the problem, because
 this will destroy very good agreement for the mass loss rate. It should be stressed here
 the good agreement in the mass loss rate between $N$-body and AGM models. This shows
 the adopted simplified model for the mass loss from the AGM describes well the
 complicated process of mass loss from the stellar systems. Finally, it
 should be stressed that the noisier appearance of the $N$-body
 model with 16000 particles as compared to that with 5000
 (see Fig. 4a) is connected with the fact that the
 larger model undergoes gravothermal like oscillations, whereas the smaller one shows
 a smooth post-collapse evolution. However, in the collapse phase, as
 expected, the smaller $N$-body model is noisier.

 \begin{figure}
 \vspace{1.0 cm}
\psfig{figure=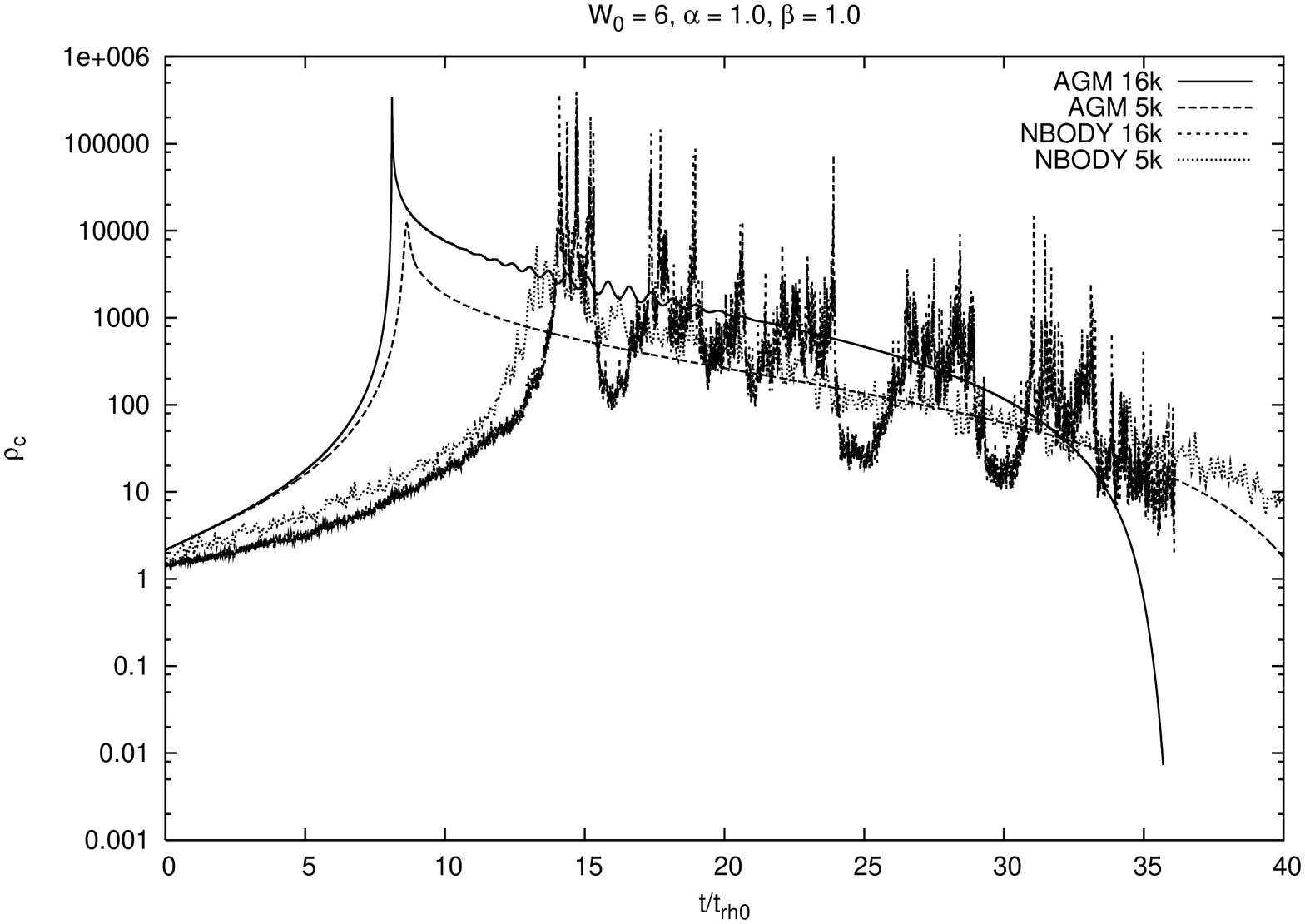,height=0.475\textwidth, width=0.475\textwidth, angle=-90}
 \caption{{\bf a:} Central density of initial King model $W_0=6$ as
 a function of time for gaseous models with 5000 (5k) and 16000 (16k) particles,
 comparing data with the corresponding direct $N$-body models, see keys.}
 \end{figure}

 \begin{figure}
 \vspace{1.0 cm}
\psfig{figure=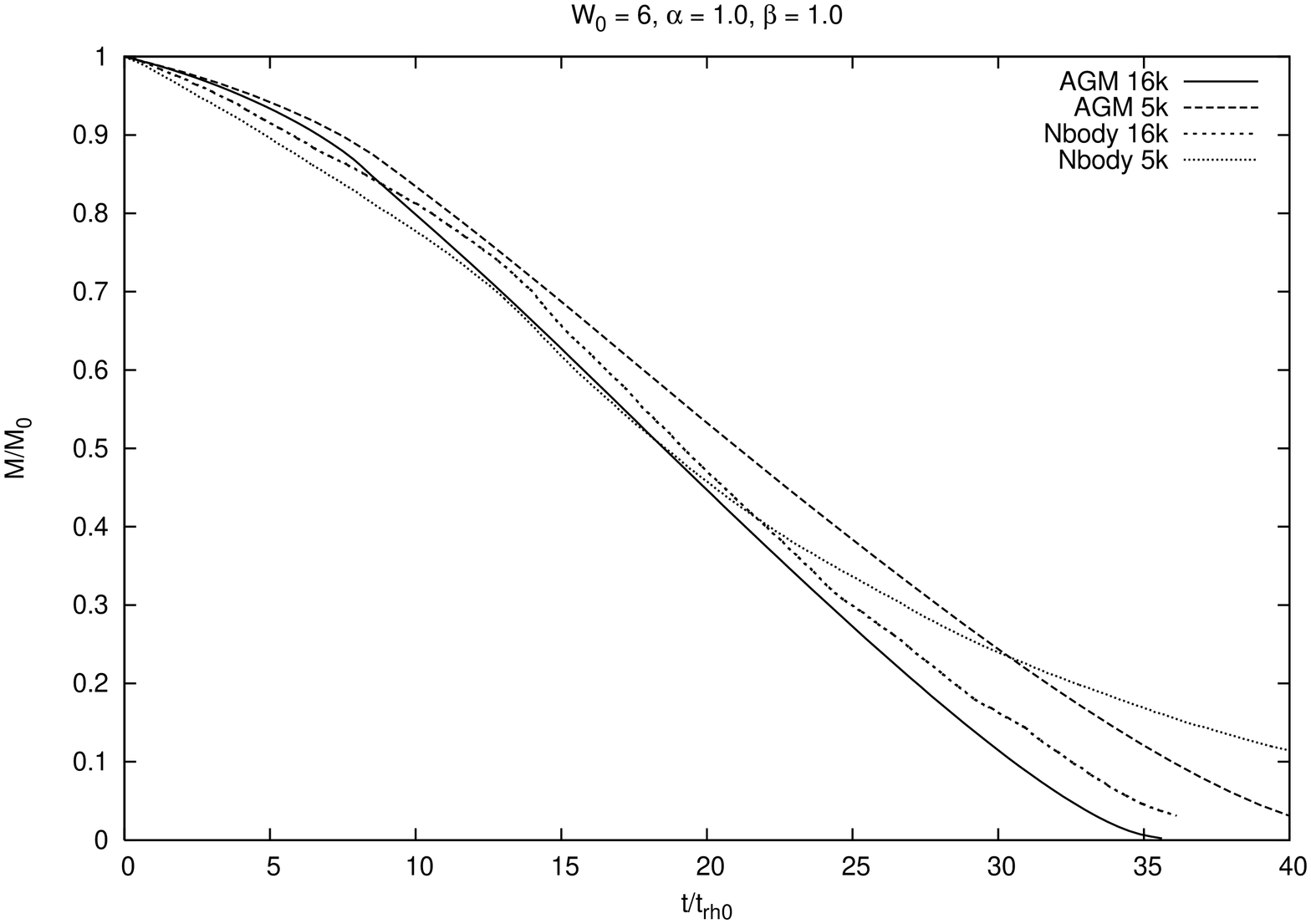,height=0.475\textwidth, width=0.475\textwidth, angle=-90}
 \contcaption{{\bf b:} As Fig. 4a, but for the total mass of the system.}
 \label{f3}
 \end{figure}

 \begin{figure}
 \vspace{1.0 cm}
\psfig{figure=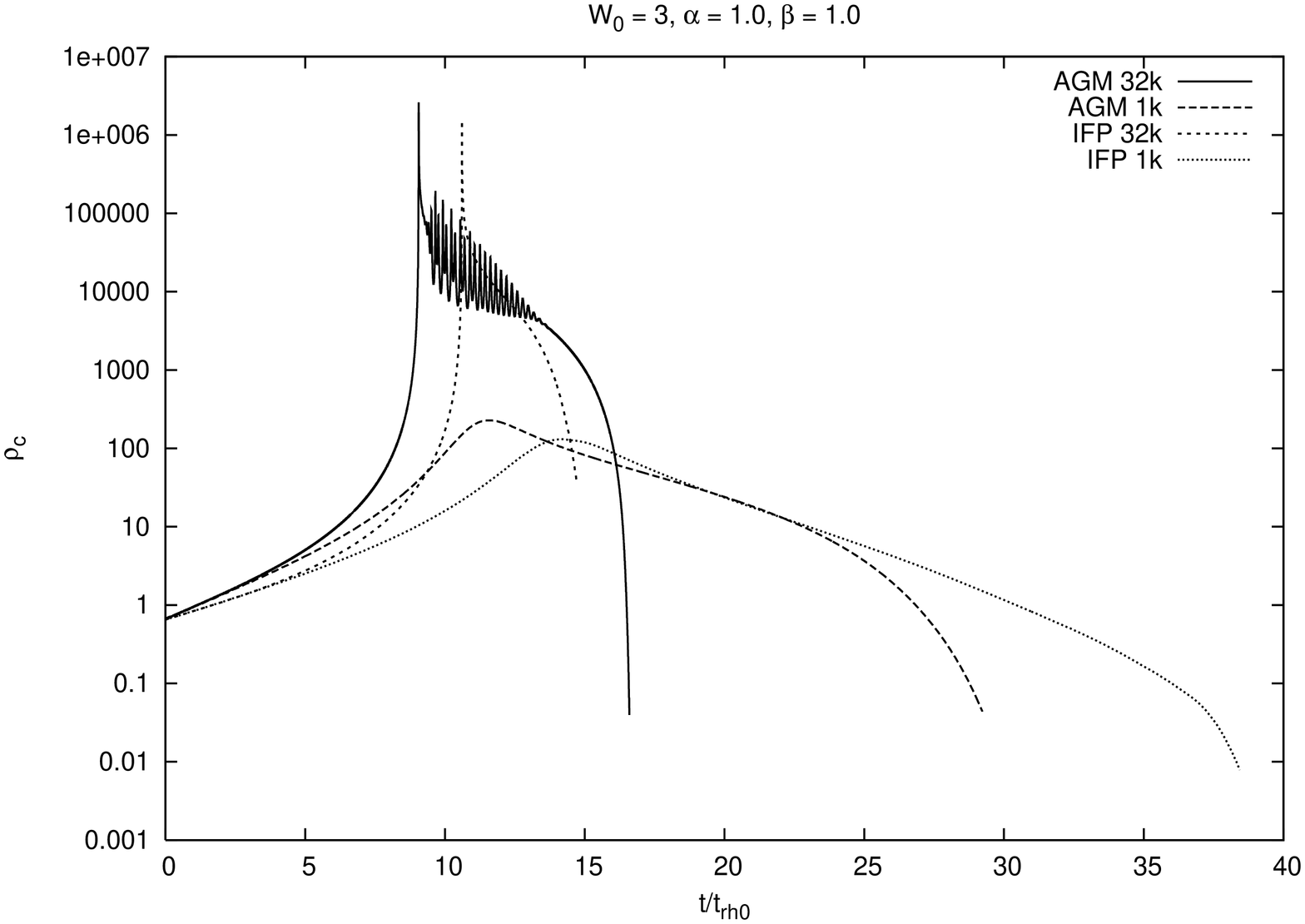,height=0.475\textwidth,
width=0.475\textwidth, angle=-90}
 \caption{{\bf a:} Central density of initial King model $W_0=3$ as
 a function of time for anisotropic gaseous (AGM) and isotropic FP (IFP) models
 ($\alpha_{\rm FP}=1.0$)
 with 1000 (1k) and 32000 (32k) particles, $\beta = 1.0$ and $\alpha = 1.0$.}
 \end{figure}

 \begin{figure}
 \vspace{1.0 cm}
\psfig{figure=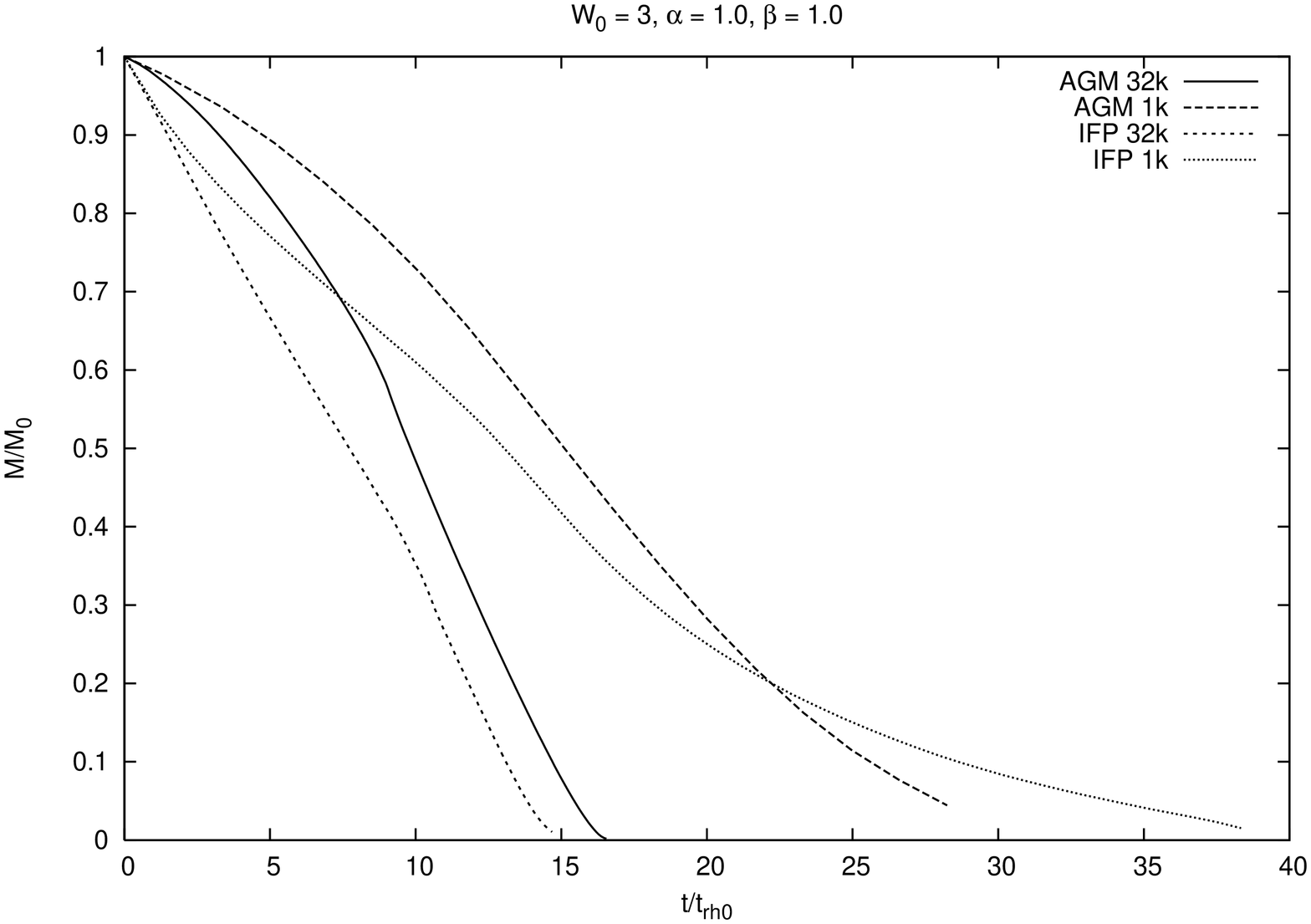,height=0.475\textwidth,
width=0.475\textwidth, angle=-90}
 \contcaption{{\bf b:} As Fig. 5a, but for the total mass of the system.}
 \label{f4}
 \end{figure}

\begin{figure}
 \vspace{1.0 cm}
\psfig{figure=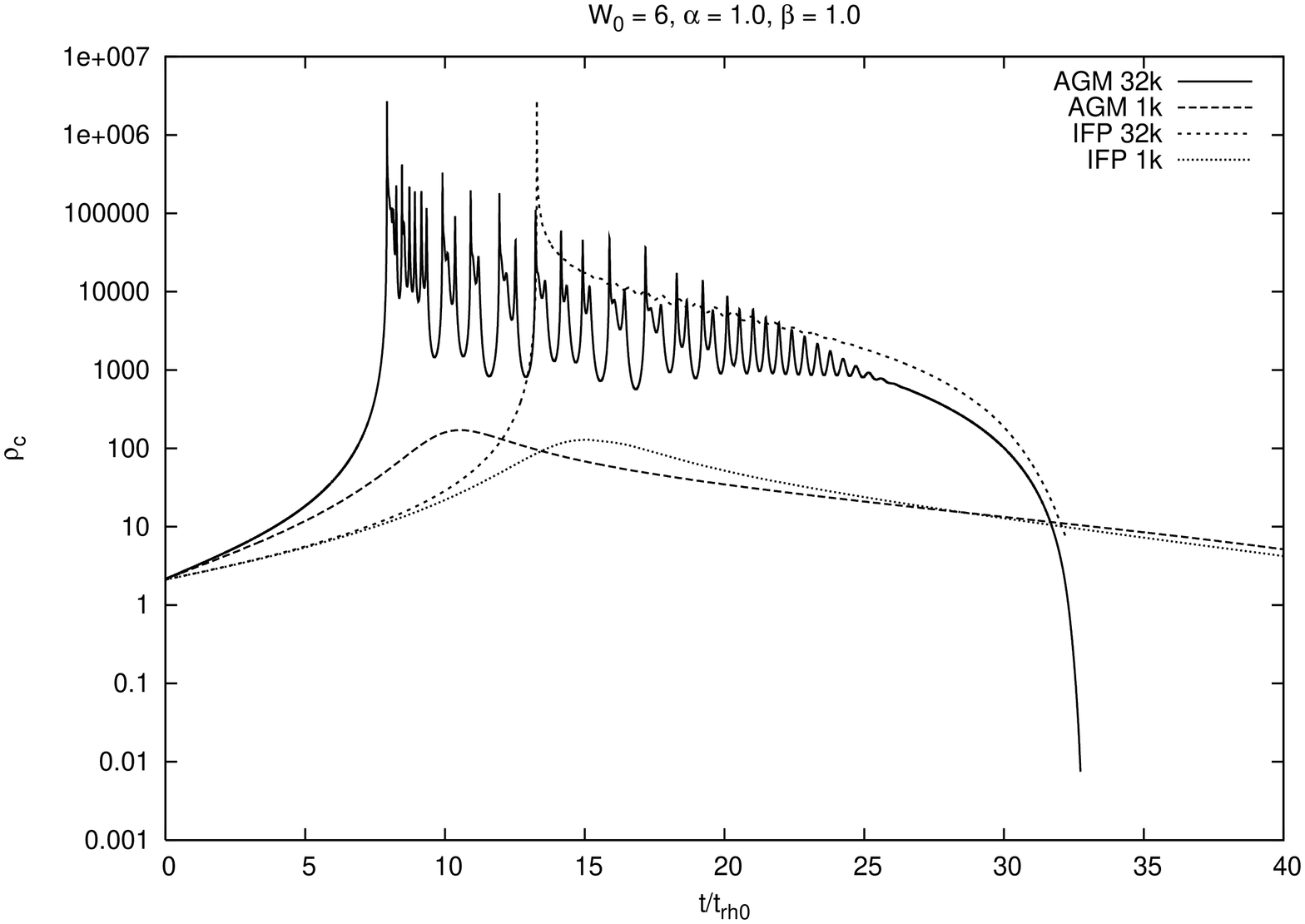,height=0.475\textwidth,
width=0.475\textwidth, angle=-90}
 \caption{{\bf a:} Central density of initial King model $W_0=6$ as
 a function of time for anisotropic gaseous (AGM) and isotropic FP ( IFP) models
 ($\alpha_{\rm FP}=1.0$)
 with 1000 (1k) and 32000 (32k) particles, $\beta = 1.0$ and $\alpha = 1.0$.}
 \end{figure}

 \begin{figure}
 \vspace{1.0 cm}
\psfig{figure=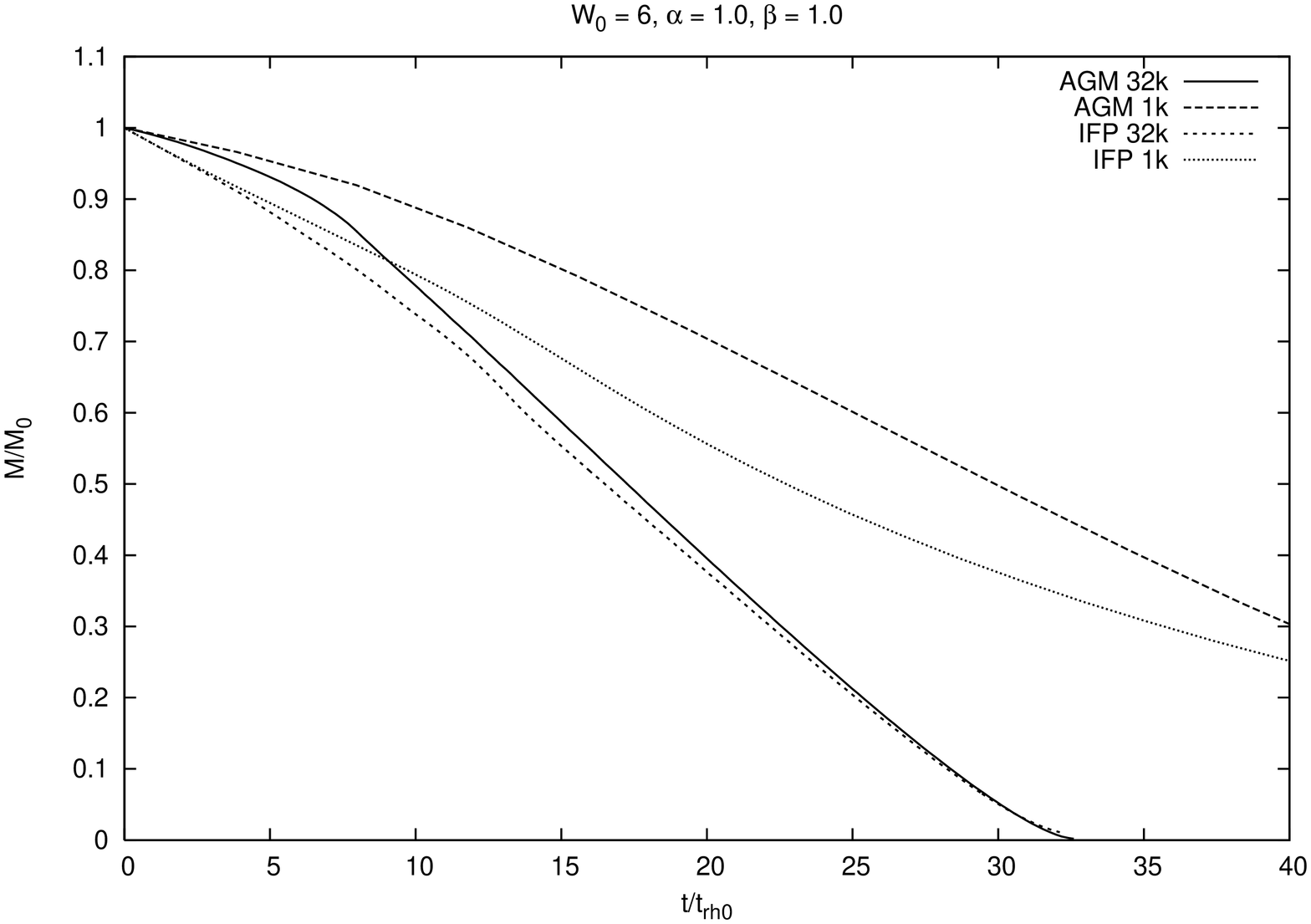,height=0.475\textwidth,
width=0.475\textwidth, angle=-90}
 \contcaption{{\bf b:} As Fig. 6a, but for the total mass of the system.}
 \label{f5}
 \end{figure}

 Now, we show a comparison of our AGM results
 with isotropic 1D Fokker-Planck results obtained ky K. Takahashi's code.
 The results are shown in Figs. 5, 6. One sees that now, the overall agreement
 between anisotropic gaseous and isotropic FP modes for different King model
 concentration parameter ($W_0 = 3, 6$) is quite good for the standard set of
 parameters: $\alpha = 1$, $\beta = 1$, $\alpha_{\rm FP} = 1$, in particular
 regarding the mass loss, but less good for the collapse time. The small adjustment of
 these parameters will give a better agreement of the mass loss rate, but still will
 not change much the picture for the collapse time. It seems that generally tidally
 limited anisotropic gaseous models intrinsically show a too fast collapse
 time. AGM predicts a collapse time of about $60 \%$ of the
 $N$-body one. It will be discussed further below in connection with
 Fig.~7 that this may be due to an incorrect distribution of tidal mass
 loss across radial shells in our model compared to the $N$-body results
 (nevertheless we get obvioulsy the total mass loss rate well). So, taking
 for example
 too much mass out of the core would change the collapse time incorrectly.
 Note that comparing individual $N$-body models (like we do
 here) with AGM creates further deviations in core collapse time
 (see for a single $N=10000$ $N$-body run Spurzem \& Aarseth 1996), because,
 as Giersz \& Heggie (1994a,b) and Giersz \& Spurzem (1994) 
 have shown, a proper ensemble average
 of direct $N$-body models makes the convergence to AGM results reliable. 
 In the cited papers we find that
 AGM predicts core collapse times in a generally very good agreement with $N$-body
 models. 

\begin{figure}
 \vspace{1.0 cm}
\psfig{figure=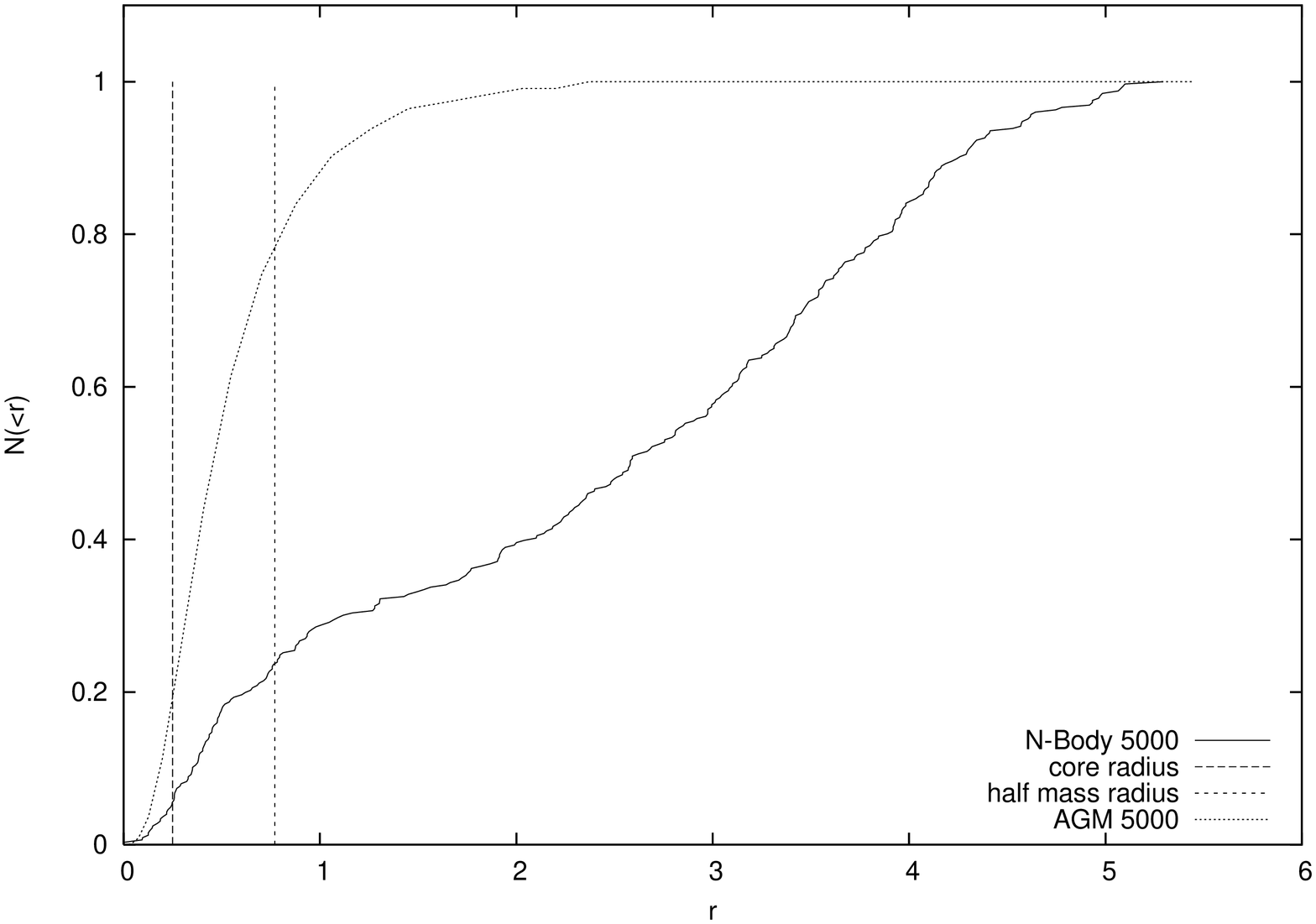,height=0.455\textwidth,
width=0.450\textwidth, angle=-90}
 \caption{Cumulative number of escapers as a function of radius, at which the last scattering occurred,
 which lifted the star to higher than escape energy. We compare for 5000 particles, close
 to the initial $W_0 = 6$ the results of AGM and direct $N$-body. Also the actual core
 and half-mass radii are indicated.}
 \label{f7}
 \end{figure}

 One might think that
 a further adjustment of the gaseous model parameters (such as the general
 conductivity $\lambda$, see e.g. Giersz \& Spurzem (1994), which scales
 the collapse time in isolated models without mass loss) would improve the situation.
 After all, the standard value of $\lambda = 0.4977$, corresponding to $C=0.104$ in
 the models of Giersz \& Heggie (1994a), has been never checked for the
 case of tidally limited systems. In a larger number of further numerical
 experiments we find, however, that there is a non-trivial relation between
 the value of $\lambda$, the core collapse time, and the value of $\beta$, which
 changes the mass loss rate, and thus also the core collapse time (see Figs. 1a,b).
 It was impossible to find a better combination of $\lambda $ and $\beta$, which
 both reproduce the mass loss and core collapse behaviour more accurately.
 We concluded that the optimal combination of parameters is the use of the standard
 value of $\lambda = 0.4977$ together with $\beta =1$; it provides good agreement
 of mass loss rates, but an error in the core collapse times has to be tolerated
 at this point. One possible reason is a still different origin of escapers in
 the AGM and the $N$-body model. Fig.~\ref{f7} shows the cumulative number of escapers
 as a function of the radius
 at which the last scattering took place lifting them to escape energy, comparing AGM and $N$-body
 for 5000 bodies, at a time close to the initial model. While both models show that
 escapers can origin from near the core or half-mass radius, it is clear that the
 AGM underestimates the rate of escapers from the halo. The larger mass loss from the core
 the faster the system evolution. We do not intend here
 to further adjust some gaseous models parameters for this effect, because this
 is beyond the scope of this paper.

\begin{figure}
 \vspace{1.0 cm}
\psfig{figure=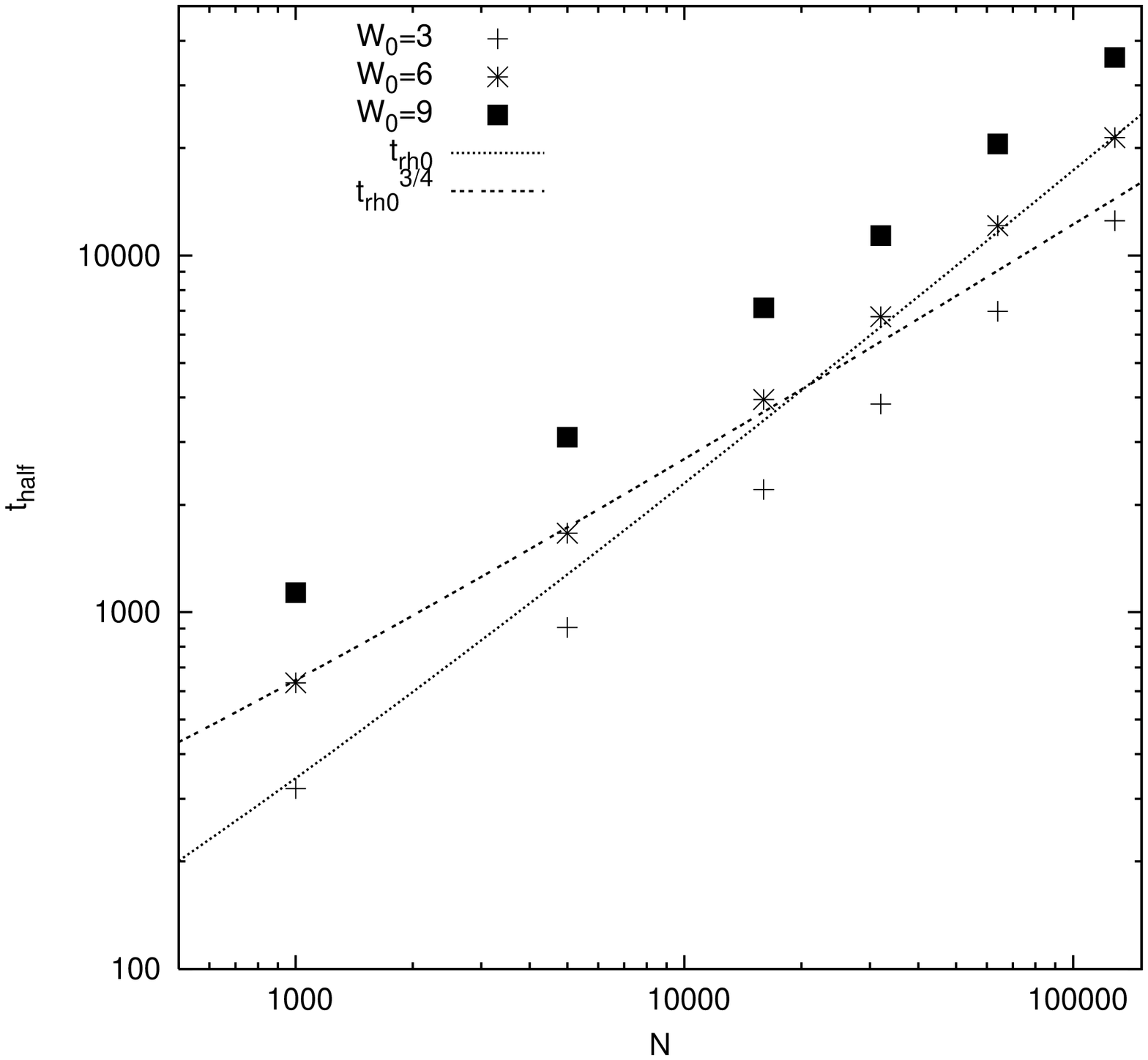,height=0.625\textwidth,
width=0.450\textwidth, angle=-90}
 \caption{Time at which the total cluster mass is equal to half of the initial mass
 for $W_0=3$, $W_0=6$ and $W_0=9$  as
 a function of initial number of particles AGM models with $\alpha = 1$,
 $\beta = 1$ and $\alpha_{\rm FP}=1.0$.}
 \label{f8}
 \end{figure}

 One can see from Fig.~\ref{f8} that the overall escape rate is very well
 modelled here in the AGM, by looking at the half-mass time (time at which the
 model will contain half of its initial mass) as a function of the initial number
 of stars in comparison to published $N$-body results.
 The scaling is proportional to $t_{rh}^{3/4}$ for low $N$; standard
 theory suggests instead a scaling proportional
 to $t_{rh}$, if potential escapers are immediately
 removed. Only for large $N$ this scaling is achieved, while for
 lower $N$ (below 50.000) it turns over to the flatter slope, which is in very good
 agreement with the theory and $N$-body results presented
 by Baumgardt (2001).

  We point out that in the
 gaseous model gravothermal oscillations in a tidally limited,
 mass-losing system are observed for the first times,
 and they are suppressed by the increasing mass
 loss towards the end of the cluster life time.
 The timestep in the FP model was chosen large
 enough to suppress oscillation for the reasons of computational time,
 so there is no conflict between the results of the two methods.
 Gaseous models are computationally cheap, even in the anisotropic
 tidally limited case. A typical model for $W_0=6$ takes about 1.5 hour
 CPU time on a Pentium 4 with 3~GHz processor (the CPU time for single mass AGM depends
 linearly on the number of shells, which in turn depends, for tidally limited
 models, only on the initial model concentration, $W_0$). So there is
 no problem in resolving the fine structure of the oscillations. Note that in
 cases where there are no post-collapse oscillations ($N=1000$) a typical
 complete run takes only several minutes of CPU time on the same computer.
 In its present form the multi-mass variant of AGM would scale in CPU
 time with $n_c^3$, where $n_c$ is the number of discrete mass components.
 This is fine for 10 components still, but becomes prohibitive at 50 or
 more components (G{\"u}rkan, Freitag \& Rasio 2004,
 but note that even in such case we are for large systems still
 much faster than direct $N$-body models). Our AGM CPU time scales
 with the third power of $n_c$ due to the
 complete implicit solution of all equations. It would
 become much faster if, as usual in FP multi-mass studies, the
 dynamical evolution of each component and the interaction via
 equipartition terms would be decoupled in the numerical solution into
 two steps. Work on this problem is in progress.

 \section{Conclusion}

 We have derived suitable model equations and their numerical
 solutions for star clusters in a tidal field with mass loss in the
 framework of the anisotropic gaseous models (AGM, Louis \& Spurzem 1991,
 Spurzem 1994, Giersz \& Spurzem 1994). The equations properly model
 the $N$-dependence of the mass loss as discussed by
 Takahashi \& Portegies Zwart (1998, 2000), Portegies Zwart \& Takahashi (1999);
 since escapers are not
 removed immediately from the system but with a time scale of
 order of the local crossing time, small systems retain
 their escapers for a longer time with respect to their
 relaxation time scale than large clusters.
 Our results reproduce the evolution of the mass loss of the
 cluster in good agreement with Fokker-Planck (FP) and direct
 $N$-body models; some discrepancy in the core collapse time
 (gaseous model collapses too fast by some 40\%) has to be accepted
 at the present time. We think that a possible reason for this
 problem is the different radial distribution of the origin of
 escapers in the clusters between AGM and direct $N$-body model,
 as seen in Fig.~\ref{f7}. An improvement of this is beyond the
 scope of this paper. Still our gaseous models provide an excellent
 tool to model the global mass loss behaviour with a very efficient
 code (see Fig.~\ref{f8}).

 Cluster evolution in tidal fields proceeds in
 three main phases according to our results. First,
 the pre-collapse evolution, which is only slightly modified
 by the mass loss as compared to the isolated model. Second,
 the steady mass loss phase in post-collapse, which in the
 framework of the gaseous model can be described by a
 simplified diffusion - escape picture, and drives
 the system steadily to smaller and smaller mass. At some
 time the mass is small enough to lift more and more stars
 across the tidal energy, and a runaway mass loss sets in,
 which is properly described by the mass loss formula
 of Eq. \ref{17}. While in the FP model the
 diffusion in energy and angular momentum is naturally
 included (which had to be added as an additional feature
 in the gaseous model), the final runaway mass loss is
 not well described by both models, even not
 in the FP model, unless it is completed by the
 term of Eq.~\ref{16}.
 In the gaseous model it is possible to switch off
 the diffusion across the tidal energy; when doing so,
 we find that the total mass evolution is extremely slow,
 practically halted, even if the mass loss term Eq. \ref{17}
 is included. Only the presence of steady mass loss caused
 by diffusion leads to conditions where the runaway
 dissolution of the cluster finally can take place. Gaseous models
 have here again shown their ability as excellent tools
 to analyze the physical processes going on in the
 evolution of heat conducting spheres, as a model for
 relaxing star clusters.

 As shown in Fig.~\ref{f8} we find that
 half-mass times ($t_{\rm half}$, time at which the
 model will contain half of its initial mass)
 scale proportional to $t_{rh}^{3/4}$. Standard theory predicts
 to $t_{\rm half} \propto t_{rh}$, if potential escapers are immediately
 removed. Our result using the AGM, however, is
 in excellent agreement with an improved theory and $N$-body results presented
 by Baumgardt (2001), who points out that there is a difference
 in time between the moment when the star reaches an energy necessary to escape and the
 moment when it leaves the cluster. Potential escapers can be scattered back to
 non-escaping energy during this time and become bound again. This process is responsible
 for a deviation from the standard scaling and leads to $t_{\rm half} \propto t_{rh}^{3/4}$
 for particle numbers at or smaller than 50000. This good agreement with $N$-body
 results additionally confirms the way in which the escapers are treated in AGM and shows
 that AGM can be used with success in simulations of tidally limited star cluster.

 For large $N$ (32000) we observe gravothermal
 oscillations (Bettwieser \& Sugimoto 1984, Goodman 1987, Makino 1996)
 during the steady mass loss phase in post-collapse.
 We observe for the first time
 the suppression of post-collapse gravothermal oscillations by
 a critically increasing mass loss at the end of the cluster life time
 (see Fig. \ref{f4}a).
 Note that earlier isotropic FP models of Drukier, Fahlman \& Richer
 (1992) and Drukier (1993) also observed cases of gravothermal oscillations
 in tidally limited clusters, but did not follow the cluster
 evolution to full dissolution.
 The choice of the timestep in the FP model shown here was large
 enough to suppress the oscillations.

 The difference between collapse times for AGM and
 $N$-body and FP models remains unsolved at the present code version.
 We did extensive parameter studies, many more than
 actually reported in this paper, to find a better optimal combination
 of parameters used to describe the mass loss in AGM ($\alpha_{\rm FP}$, $\alpha$, $\beta$),
 sometimes even also varying the standard scaling parameter for the conductivity
 $\lambda \equiv 0.4977$ (in Heggie's models it is $C \equiv 27\sqrt{\pi}\lambda/10$, see e.g.
 Giersz \& Spurzem (1994). The result is that the standard parameters
 give the best agreement for all other properties between $N$-body, FP and
 AGM models, just the core collapse time remains too short in AGM.
 It should be noted that also the agreement
 between FP models and $N$-body models is not perfect regarding
 the collapse time. We speculate that the radial origin of escapers
 as described above may be the main reason for the discrepancy.

 Note that our anisotropic gaseous models (AGM) have been used in
 their multi-mass form already with 50 components in G{\"u}rkan,
 Freitag, \& Rasio (2004) and in Boily et al. (2005) with 15 
 components for studies of mass segregation.
 We want to study the future our models with the improved
 tidal boundary in connection with more realistic systems with a stellar
 mass spectrum (cf. e.g. the collaborative experiment,
 Heggie et al. 1998) and stellar evolution, and do a further comparison 
 with a set of $N$-body models, possibly improving also the remaining
 problems in AGM.

\section*{Appendix}

For clarity we have defined the following integrals in analogy to the
error function (showed first):
\begin{eqnarray*}
\erf(x) &=& \twosqpi \intl_0^x \exp(-t^2) dt  \cr
\Erf(x) &=& \twosqpi \intl_0^x t^2  \exp(-t^2) dt  \cr
\I(x) &=& \twosqpi \intl_0^x \exp(t^2) dt  \cr
\J(x) &=& \twosqpi \intl_0^x t^2  \exp(t^2) dt
\end{eqnarray*}
Note the relations
\begin{eqnarray*}
\Erf(x) &=& {1\over 2} \erf(x) - {x\over \sqrt{\pi}} \exp(-x^2) \cr
\J(x) &=& -{1\over 2} \I(x) + {x\over \sqrt{\pi}} \exp(x^2)  \cr
\I(x) &=& \twosqpi \exp(x^2) \D(x)
\end{eqnarray*}
where $\D(x) $ is Dawson's Integral defined as
\begin{eqnarray*}
\D(x) &=& \exp(-x^2) \intl_0^x t^2  \exp(t^2) dt  \hfill
\end{eqnarray*}
While for the standard error function we use the intrinsic function
provided by the standard fortran compilers, Dawson's integral
has to be taken from the Numerical Recipes, Chapter 6.10 (Press et al 1986).
Dawson's Integral vanishes for $x\rightarrow 0$. Since the
standard method given in the Recipes involves exponential
functions this is not good for small argument values, therefore
for  $|x|<0.2$ Dawson's Integral is in the given recipe evaluated by a
Taylor series up to order $x^7$.
In our expressions for $X_e$, $X_r$, and $X_t$ we have similarly ill-behaved
functions, namely $\erf(x)/x$, $\Erf(x)/x^3$, $\I(x)/x$, and $\J(x)/x^3$.
All these expressions have to be taken for the arguments $G \equiv \sqrt{b^2-a^2}, b>a$
or $H \equiv \sqrt{a^2-b^2}, a>b$, and they approach zero for $b\rightarrow a$.
In our numerical computation of these functions we therefore also
use the following series expansions for $|x|<0.2$:
\begin{eqnarray*}
{\I(x)\over x},{\erf(x)\over x} &\approx &
  \twosqpi\Bigl(1 \pm {1\over 3} x^2
   \bigl( 1 \pm {3\over 10} x^2 ( 1 \pm {5\over 21}x^2)\bigr)\Bigr)\cr
{\J(x)\over x^3},{\Erf(x)\over x^3} &\approx &
  \twosqpi\Bigl({1\over 3} \pm {1\over 5} x^2
   \bigl( 1 \pm {5\over 14} x^2 ( 1 \pm  {7\over 27}x^2)\bigr)\Bigr) \cr
\end{eqnarray*}
The $\pm$ sign should be taken as a $+$ for $\I$, $\J$ (involving $\exp(t^2)$) and as a $-$ for
$\erf$, $\Erf$ (involving $\exp(-t^2)$).

While the above Taylor series are completely well behaved for $x\rightarrow 0$, it
is nevertheless instructive to look at the asymptotic forms for $X_e$, $X_r$, and
$X_t$ which are obtained from the following asymptotic equalities for $x\rightarrow 0$:
\begin{eqnarray*}
{\I(x)\over x},{\erf(x)\over x} &\sim & \twosqpi \exp(\pm x^2) \cr
{\J(x)\over x^3},{\Erf(x)\over x^3} &\sim &  \twosqpi {1\over 3}  \exp(\pm x^2)
\end{eqnarray*}
The use of $\pm$ is to be understood as above. With $b=a$ we get the results
\begin{eqnarray*}
X_e &=& \erf(a) - \twosqpi \cdot a \cdot\exp(-a^2) \cr
X_r &=& \erf(a) - \twosqpi \cdot a (1\!+\!{2\over 3} a^2)\cdot \exp(-a^2) \cr
X_t &=& \erf(a) - \twosqpi \cdot a (1\!+\!{2\over 3} a^2)\cdot \exp(-a^2) = X_r \cr
\end{eqnarray*}
This same result is obtained approaching $b=a$ from both sides ($b<a$, $b>a$),
and it reaffirms that in the case of an isotropic velocity distribution and
equal escape velocities in both the radial and tangential direction (i.e. using
an energy criterion for escape) we have isotropy for the energy of the
escaping stars.

 \section*{Acknowledgements}
 Part of this work was performed in the context of
 German Science Foundation (DFG) grant Sp 345/10-1,2.
KT was supported by the Research for the Future Program of Japan
Society for the Promotion of Science (JSPS-RFTP97P01102). MG was
supported by the Polish State Committee for Science Research (KBN)
grant 0394/P03/2004/27. RS and MG acknowledge support from
the DFG Central and Eastern Europe support grant 436 POL 113/103.
Numerical simulations were done partly at the
IBM Jump supercomputer of the
John von Neumann-Institute for Computing (NIC) J"ulich, Germany.
We thank Douglas Heggie, Stefan Deiters, Holger Baumgardt,
Pau Amaro-Seoane and Marc Freitag for
very helpful discussions and suggestions and
an anonymous referee for a careful and helpful report.


\begin{thebibliography}{}
\bibitem[]{} Aarseth S.J., 1985, in Brackbill J.U.,
   Cohen B.I., eds, Multiple time scales, Academic Press, Orlando,
      p. 378
\bibitem[\protect\citeauthoryear{Aarseth}{1999}]{1999PASP..111.1333A}
Aarseth S.~J., 1999a, PASP, 111, 1333

\bibitem[\protect\citeauthoryear{Aarseth}{1999}]{1999CeMDA..73..127A}
Aarseth S.~J., 1999b, CeMDA, 73, 127

 \bibitem[\protect\citeauthoryear{Aarseth}{2003}]{2003aar}
Aarseth S.J., 2003, Gravitational N-Body Simulations:
Tools and Algorithms (Cambridge Monographs on Mathematical Physics)

\bibitem[\protect\citeauthoryear{Aarseth \&
Heggie}{1998}]{1998MNRAS.297..794A} Aarseth S.~J., Heggie D.~C., 1998,
MNRAS, 297, 794

\bibitem[\protect\citeauthoryear{Amaro-Seoane, Freitag, \&
Spurzem}{2004}]{2004MNRAS.352..655A} Amaro-Seoane P., Freitag M., Spurzem
R., 2004, MNRAS, 352, 655

\bibitem[\protect\citeauthoryear{Baumgardt}{2001}]{2001MNRAS.325.1323B}
Baumgardt H., 2001, MNRAS, 325, 1323

\bibitem[\protect\citeauthoryear{Bettwieser}{1983}]{1983MNRAS.203..811B} 
Bettwieser E., 1983, MNRAS, 203, 811 
 
\bibitem[\protect\citeauthoryear{Bettwieser \&
Sugimoto}{1984}]{1984MNRAS.208..493B} Bettwieser E., Sugimoto D., 1984,
MNRAS, 208, 493

\bibitem[\protect\citeauthoryear{Boily et al.}{2005}]{2005ApJ...620L..27B} 
Boily C.~M., Lan{\c c}on A., Deiters S., Heggie D.~C., 2005, ApJ, 620, L27 

\bibitem[\protect\citeauthoryear{Chernoff \&
Weinberg}{1990}]{1990ApJ...351..121C} Chernoff D.~F., Weinberg M.~D., 1990,
ApJ, 351, 121

\bibitem[\protect\citeauthoryear{Cohn}{1979}]{1979ApJ...234.1036C} Cohn H.,
1979, ApJ, 234, 1036

\bibitem[\protect\citeauthoryear{Cohn}{1980}]{1980ApJ...242..765C} Cohn H.,
1980, ApJ, 242, 765

\bibitem[\protect\citeauthoryear{Drukier}{1993}]{1993MNRAS.265..773D}
Drukier G.~A., 1993, MNRAS, 265, 773

\bibitem[\protect\citeauthoryear{Drukier et
al.}{1999}]{1999ApJ...518..233D} Drukier G.~A., Cohn H.~N., Lugger P.~M.,
Yong H., 1999, ApJ, 518, 233

\bibitem[\protect\citeauthoryear{Drukier, Fahlman, \&
Richer}{1992}]{1992ApJ...386..106D} Drukier G.~A., Fahlman G.~G., Richer
H.~B., 1992, ApJ, 386, 106

\bibitem[\protect\citeauthoryear{Einsel \&
Spurzem}{1999}]{1999MNRAS.302...81E} Einsel C., Spurzem R., 1999, MNRAS,
302, 81

\bibitem[\protect\citeauthoryear{Fiestas, Kim \& Spurzem}{2005}]{2005MNRAS.....}
Fiestas J., Spurzem R., Kim E., 2005, MNRAS, subm.

\bibitem[\protect\citeauthoryear{Freitag}{2000}]{frei2000}
Freitag M., Ph.D. Thesis 2000, Univ. de Gen\`eve, Switzerland

\bibitem[\protect\citeauthoryear{Freitag \&
Benz}{2001}]{2001A&A...375..711F} Freitag M., Benz W., 2001, A\&A, 375, 711

\bibitem[\protect\citeauthoryear{Freitag \& Benz}{2002}]{2002A&A...394..345F}
Freitag M., Benz W., 2002, A\&A, 394, 345

\bibitem[\protect\citeauthoryear{Frank \& Rees}{1976}]{1976MNRAS.176..633F}
Frank J., Rees M.~J., 1976, MNRAS, 176, 633

\bibitem[\protect\citeauthoryear{Fregeau et
al.}{2003}]{2003ApJ...593..772F} Fregeau J.~M., G{\" u}rkan M.~A., Joshi
K.~J., Rasio F.~A., 2003, ApJ, 593, 772

\bibitem[\protect\citeauthoryear{Fukushige \&
Heggie}{2000}]{2000MNRAS.318..753F} Fukushige T., Heggie D.~C., 2000,
MNRAS, 318, 753

\bibitem[\protect\citeauthoryear{Giersz}{1996}]{1996IAUS..174..101G} Giersz
M., 1996, IAUS, 174, 101

\bibitem[\protect\citeauthoryear{Giersz}{1998}]{1998MNRAS.298.1239G} Giersz
M., 1998, MNRAS, 298, 1239

\bibitem[\protect\citeauthoryear{Giersz}{2001}]{2001MNRAS.324..218G} Giersz
M., 2001, MNRAS, 324, 218

\bibitem[\protect\citeauthoryear{Giersz \&
Heggie}{1994a}]{1994MNRAS.268..257G} Giersz M., Heggie D.~C., 1994a,
MNRAS, 268, 257

\bibitem[\protect\citeauthoryear{Giersz \&
Heggie}{1994b}]{1994MNRAS.270..298G} Giersz M., Heggie D.~C., 1994b,
MNRAS, 270, 298

\bibitem[\protect\citeauthoryear{Giersz \&
Heggie}{1996}]{1996MNRAS.279.1037G} Giersz M., Heggie D.~C., 1996, MNRAS,
279, 1037

\bibitem[\protect\citeauthoryear{Giersz \&
Spurzem}{1994}]{1994MNRAS.269..241G} Giersz M., Spurzem R., 1994, MNRAS,
269, 241

\bibitem[\protect\citeauthoryear{Giersz \&
Spurzem}{2000}]{2000MNRAS.317..581G} Giersz M., Spurzem R., 2000, MNRAS,
317, 581

\bibitem[\protect\citeauthoryear{Giersz \&
Spurzem}{2003}]{2003MNRAS.343..781G} Giersz M., Spurzem R., 2003, MNRAS,
343, 781

\bibitem[\protect\citeauthoryear{Goodman}{1987}]{1987ApJ...313..576G}
Goodman J., 1987, ApJ, 313, 576

\bibitem[\protect\citeauthoryear{Grillmair et
al.}{1999}]{1999AJ....117..167G} Grillmair C.~J., Forbes D.~A., Brodie
J.~P., Elson R.~A.~W., 1999, AJ, 117, 167

\bibitem[\protect\citeauthoryear{G{\" u}rkan, Freitag, \&
Rasio}{2004}]{2004ApJ...604..632G} G{\" u}rkan M.~A., Freitag M., Rasio
F.~A., 2004, ApJ, 604, 632

\bibitem[\protect\citeauthoryear{Hansen et al.}{2002}]{2002ApJ...574L.155H}
Hansen B.~M.~S., et al., 2002, ApJ, 574, L155

\bibitem[\protect\citeauthoryear{Heggie}{1984}]{1984MNRAS.206..179H} Heggie
D.~C., 1984, MNRAS, 206, 179

\bibitem[\protect\citeauthoryear{Heggie \&
Aarseth}{1992}]{1992MNRAS.257..513H} Heggie D.~C., Aarseth S.~J., 1992,
MNRAS, 257, 513

\bibitem[\protect\citeauthoryear{Heggie et al.}{1998}]{1998HiA....11..591H}
Heggie D.~C., Giersz M., Spurzem R., Takahashi K., 1998,  J. Andersen, ed,
 Highlights of Astronomy Vol. 11, Kluwer Acad. Publishers, p. 591

\bibitem[\protect\citeauthoryear{Heggie \&
Mathieu}{1986}]{1986LNP...267..233H} Heggie D.~C., Mathieu R.~D.,
1986, in Hut P., McMillan S.L.W., eds., The Use of Supercomputers in
Stellar Dynamics., Springer Berlin, Lect. Notes in Physics, 267, 233

\bibitem[\protect\citeauthoryear{Hut \& Makino}{1999}]{1999Sci...283..501H}
Hut P., Makino J., 1999, Science, 283, 501

\bibitem[\protect\citeauthoryear{Ibata et al.}{1999}]{1999ApJS..120..265I}
Ibata R.~A., Richer H.~B., Fahlman G.~G., Bolte M., Bond H.~E., Hesser
J.~E., Pryor C., Stetson P.~B., 1999, ApJS, 120, 265

\bibitem[\protect\citeauthoryear{Joshi, Rasio, \& Portegies
Zwart}{2000}]{2000ApJ...540..969J} Joshi K.~J., Rasio F.~A., Portegies
Zwart S., 2000, ApJ, 540, 969

\bibitem[\protect\citeauthoryear{Joshi, Nave, \&
Rasio}{2001}]{2001ApJ...550..691J} Joshi K.~J., Nave C.~P., Rasio F.~A.,
2001, ApJ, 550, 691

\bibitem[\protect\citeauthoryear{Kim }{2003}]{Kim2003}
Kim E., 2003, Dynamical Evolution of Rotating Star Clusters, PhD Thesis,
 Department of Astronomy, Graduate School, Seoul National University

\bibitem[\protect\citeauthoryear{Kim et al.}{2002}]{2002MNRAS.334..310K}
Kim E., Einsel C., Lee H.~M., Spurzem R., Lee M.~G., 2002, MNRAS, 334, 310

\bibitem[\protect\citeauthoryear{Kim, Lee, \&
Spurzem}{2004}]{2004MNRAS.351..220K} Kim E., Lee H.~M., Spurzem R., 2004,
MNRAS, 351, 220

\bibitem[\protect\citeauthoryear{King}{1966}]{1966AJ.....71..276K} King
I.~R., 1966, AJ, 71, 276

\bibitem[\protect\citeauthoryear{King et al.}{1998}]{1998ApJ...492L..37K}
King I.~R., Anderson J., Cool A.~M., Piotto G., 1998, ApJ, 492, L37

\bibitem[\protect\citeauthoryear{Klessen \&
Kroupa}{1998}]{1998ApJ...498..143K} Klessen R.~S., Kroupa P., 1998, ApJ,
498, 143

\bibitem[\protect\citeauthoryear{Koch et al.}{2004}]{2004AJ....128.2274K}
Koch A., Grebel E.~K., Odenkirchen M., Mart{\'{\i}}nez-Delgado D., Caldwell
J.~A.~R., 2004, AJ, 128, 2274

\bibitem[\protect\citeauthoryear{Kroupa}{1998}]{1998MNRAS.300..200K} Kroupa
P., 1998, MNRAS, 300, 200

\bibitem[\protect\citeauthoryear{Lee \&
Ostriker}{1987}]{1987ApJ...322..123L} Lee H.~M., Ostriker J.~P., 1987, ApJ,
322, L123

\bibitem[\protect\citeauthoryear{Louis}{1990}]{1990MNRAS.244..478L} Louis 
P.~D., 1990, MNRAS, 244, 478 
 
\bibitem[\protect\citeauthoryear{Louis \&
Spurzem}{1991}]{1991MNRAS.251..408L} Louis P.~D., Spurzem R., 1991, MNRAS,
251, 408

\bibitem[\protect\citeauthoryear{Lupton, Gunn, \&
Griffin}{1987}]{1987AJ.....93.1114L} Lupton R.~H., Gunn J.~E., Griffin
R.~F., 1987, AJ, 93, 1114

\bibitem[\protect\citeauthoryear{Lynden-Bell \& 
Eggleton}{1980}]{1980MNRAS.191..483L} Lynden-Bell D., Eggleton P.~P., 1980, 
MNRAS, 191, 483 

\bibitem[\protect\citeauthoryear{Makino}{1996}]{1996ApJ...471..796M} Makino
J., 1996, ApJ, 471, 796

\bibitem[\protect\citeauthoryear{Makino \&
Aarseth}{1992}]{1992PASJ...44..141M} Makino J., Aarseth S.~J., 1992, PASJ,
44, 141

\bibitem[\protect\citeauthoryear{Makino \& Hut}{1988}]{1988ApJS...68..833M}
Makino J., Hut P., 1988, ApJS, 68, 833

\bibitem[\protect\citeauthoryear{Makino \&
Taiji}{1998}]{1998sssp.book.....M} Makino J., Taiji M., 1998,
Scientific simulations with special-purpose computers : The GRAPE systems,
Chichester ; Toronto : John Wiley \& Sons.

\bibitem[\protect\citeauthoryear{Makino et al.}{1997}]{1997ApJ...480..432M}
Makino J., Taiji M., Ebisuzaki T., Sugimoto D., 1997, ApJ, 480, 432

\bibitem[\protect\citeauthoryear{Odenkirchen et
al.}{2001}]{2001ApJ...548L.165O} Odenkirchen M., et al., 2001, ApJ, 548,
L165

\bibitem[\protect\citeauthoryear{Piotto \&
Zoccali}{1999}]{1999A&A...345..485P} Piotto G., Zoccali M., 1999, A\&A,
345, 485

\bibitem[\protect\citeauthoryear{Piotto et al.}{1999}]{1999AJ....117..264P}
Piotto G., Zoccali M., King I.~R., Djorgovski S.~G., Sosin C., Dorman B.,
Rich R.~M., Meylan G., 1999, AJ, 117, 264

\bibitem[\protect\citeauthoryear{Portegies Zwart \&
Takahashi}{1999}]{1999CeMDA..73..179P} Portegies Zwart S.~F., Takahashi K.,
1999, CeMDA, 73, 179

\bibitem[\protect\citeauthoryear{Press et al.}{1986}]{} Press W.H., Flannery B.P., Teukolsky S.A., Vetterling W.T., 1986, Numerical Recipes. Cambridge University Press, Cambridge

\bibitem[\protect\citeauthoryear{Quinlan}{1996}]{1996NewA....1..255Q}
Quinlan G.~D., 1996, NewA, 1, 255

\bibitem[\protect\citeauthoryear{Richer et al.}{2002}]{2002ApJ...574L.151R}
Richer H.~B., et al., 2002, ApJ, 574, L151

\bibitem[\protect\citeauthoryear{Rosenbluth, MacDonald, \& 
Judd}{1957}]{1957PhRv..107....1R} Rosenbluth M.~N., MacDonald W.~M., Judd 
D.~L., 1957, PhRv, 107, 1 

\bibitem[\protect\citeauthoryear{Rubenstein \&
Bailyn}{1999}]{1999ApJ...513L..33R} Rubenstein E.~P., Bailyn C.~D., 1999,
ApJ, 513, L33

\bibitem[\protect\citeauthoryear{Shara et al.}{1998}]{1998ApJ...508..570S}
Shara M.~M., Fall S.~M., Rich R.~M., Zurek D., 1998, ApJ, 508, 570

\bibitem[\protect\citeauthoryear{Spitzer}{1987}]{1987degc.book.....S}
Spitzer L., 1987,  Dynamical Evolution of Globular Clusters.
Princeton Univ. Press, Princeton

\bibitem[\protect\citeauthoryear{Spurzem}{1994}]{1994LNP...430..170S}
Spurzem R., 1994, in Pfenniger D., Gurzadyan V.G., eds,
 Ergodic Concepts in Stellar Dynamics, Springer-Vlg.,
 Berlin, Heidelberg, Lect. Notes in Phyics, 430, 170

\bibitem[\protect\citeauthoryear{Spurzem}{1996}]{1996IAUS..174..111S}
Spurzem R., 1996, in Hut P., Makino J., eds, Dynamics
 of Star Clusters, Proc. IAU Symp. No. 174, p. 111

\bibitem[\protect\citeauthoryear{Spurzem}{1999}]{1999JCoAM.109..407S}
Spurzem R., 1999, Jl. Comp. Appl. Maths., 109, 407

\bibitem[\protect\citeauthoryear{Spurzem \&
Aarseth}{1996}]{1996MNRAS.282...19S} Spurzem R., Aarseth S.~J., 1996,
MNRAS, 282, 19

\bibitem[\protect\citeauthoryear{Spurzem \&
Giersz}{1996}]{1996MNRAS.283..805S} Spurzem R., Giersz M., 1996, MNRAS,
283, 805

\bibitem[\protect\citeauthoryear{Spurzem \&
Takahashi}{1995}]{1995MNRAS.272..772S} Spurzem R., Takahashi K., 1995,
MNRAS, 272, 772

\bibitem[\protect\citeauthoryear{Sugimoto et
al.}{1990}]{1990Natur.345...33S} Sugimoto D., Chikada Y., Makino J., Ito
T., Ebisuzaki T., Umemura M., 1990, Nature, 345, 33

\bibitem[\protect\citeauthoryear{Takahashi}{1995}]{1995PASJ...47..561T}
Takahashi K., 1995, PASJ, 47, 561

\bibitem[\protect\citeauthoryear{Takahashi}{1996}]{1996PASJ...48..691T}
Takahashi K., 1996, PASJ, 48, 691

\bibitem[\protect\citeauthoryear{Takahashi}{1997}]{1997PASJ...49..547T}
Takahashi K., 1997, PASJ, 49, 547

\bibitem[\protect\citeauthoryear{Takahashi, Lee, \&
Inagaki}{1997}]{1997MNRAS.292..331T} Takahashi K., Lee H.~M., Inagaki S.,
1997, MNRAS, 292, 331

\bibitem[\protect\citeauthoryear{Takahashi \& Portegies
Zwart}{1998}]{1998ApJ...503L..49T} Takahashi K., Portegies Zwart S.~F.,
1998, ApJ, 503, L49

\bibitem[\protect\citeauthoryear{Takahashi \& Portegies
Zwart}{2000}]{2000ApJ...535..759T} Takahashi K., Portegies Zwart S.~F.,
2000, ApJ, 535, 759

\bibitem[\protect\citeauthoryear{Takahashi \&
Lee}{2000}]{2000MNRAS.316..671T} Takahashi K., Lee H.~M., 2000, MNRAS, 316,
671

\bibitem[\protect\citeauthoryear{Vicary}{1997}]{Vicar1997}
Vicary, D.C., 1997, Master Thesis, Univ. of Edinburgh,
     Dept. of Maths. and Stats.

\bibitem[\protect\citeauthoryear{Watters, Joshi, \&
Rasio}{2000}]{2000ApJ...539..331W} Watters W.~A., Joshi K.~J., Rasio F.~A.,
2000, ApJ, 539, 331

 \end{thebibliography}
 \end{document}